\documentclass{article}

\newcommand{\bt}{\begin{tabular}{c}}
\newcommand{\et}{\end{tabular}}

\newcommand{\eb}{\ee\be } 
\newcommand{\ebp}{\rt.\ee\be\lt.} 
\newcommand{\bmat}{\lt ( \begin{array} }
\newcommand{\emat}{  \end{array} \rt )}

\newcommand{\ovv}{{\ov v}}

\newcommand{\ED}{\end{document}}

\newcommand{\og}{{\ov g}}
\newcommand{\oy}{{\ov \y}}

\newcommand{\oC}{{\ov C}}

\newcommand{\oF}{{\ov F}}
\newcommand{\A}{{\ov A}}

\renewcommand{\a}{\alpha}	
\renewcommand{\b}{\beta}
\newcommand{\g}{\gamma}
\renewcommand{\d}{\delta}
\newcommand{\e}{\epsilon}
\newcommand{\ve}{\varepsilon}

\newcommand{\m}{\mu}

\renewcommand{\r}{\rho}

\newcommand{\f}{\phi}

\newcommand{\y}{\psi}
\newcommand{\w}{\omega}

\newcommand{\D}{\Delta}
	
\renewcommand{\L}{\Lambda}

\newcommand{\F}{\Phi}

\newcommand{\la}{\label}
\newcommand{\ci}{\cite}

\newcommand{\ds}{\documentstyle}	
\newcommand{\fr}{\frac}

\newcommand{\pa}{\partial}
\newcommand{\ov}{\overline}
\newcommand{\be}{\begin{equation}}
\newcommand{\ee}{\end{equation}}
\newcommand{\ba}{\begin{array}} 
\newcommand{\ea}{\end{array}}
\newcommand{\bea}{\begin{eqnarray}}
\newcommand{\eea}{\end{eqnarray}}
\newcommand{\ra}{\rightarrow}
\newcommand{\Ra}{\Rightarrow}

\newcommand{\lt}{\left}
\newcommand{\rt}{\right}

\newcommand{\ben}{\begin{enumerate}}
\newcommand{\een}{\end{enumerate}}
\newcommand{\bitem}{\begin{itemize}}
\newcommand{\eitem}{\end{itemize}}
\newcommand{\articlenumber}{}

\newcommand{\articletitle}{ 
Detailed Calculations of the Mass Spectrum
 for the Leptons \\ after Supersymmetry Breaking \\
in the Supersymmetric Standard Model:\\ Cybersusy IV}

\begin{document}
\makeatletter	   
\renewcommand{\ps@plain}{%
\renewcommand{\@oddhead}{{\articlenumber  \hspace{1cm} }\hspace{1cm}  \hfil\textrm{\thepage}} 
\renewcommand{\@evenhead}{\@oddhead}
\renewcommand{\@oddfoot}{\textrm{\articlenumber \hspace{1cm}  }\hspace{1cm} \hfil\textrm{\thepage}}
\renewcommand{\@evenfoot}{\@oddfoot}}
\makeatother    
\title{  \articletitle \\ \articlenumber}
\author{ J. A.  Dixon\footnote{jadix@telus.net}\\ Dixon Law Firm\footnote{Fax: (403) 266-1487} \\1020 Canadian Centre\\
833 - 4th Ave. S. W. \\ Calgary, Alberta \\ Canada T2P 3T5 }
\maketitle
\pagestyle{plain}
\Large

\abstract{This is the fourth of a series of four papers introducing cybersusy, which is a new approach to supersymmetry breaking in the supersymmetric standard model. This paper contains a brief summary, and then goes on to the calculation of the propagators and masses for the leptons, using the cybersusy action deduced in the three previous papers of this series.  The results here are for the general case of three flavours of leptons.  }

\Large

\section{Introduction}

In the first paper of this series \ci{cybersusyI}, it was claimed that {\bf  supersymmetry breaks when gauge symmetry breaks}. 
This was illustrated in detail using  an effective action for leptons based on the cybersusy algebra and action in \ci{cybersusyI}. The cybersusy algebra was derived from the BRS cohomology of certain composite operators in the supersymmetric standard model (SSM). The nature of the mechanism is such that there is a separate mass spectrum for each set of conserved quantum numbers.
\ben
\item
The three electron flavours in the SSM give rise to
\ben
\item
 nine Dirac fermionic charged leptons,
\item
 nine scalar boson charged leptons, and 
\item
three vector boson charged leptons
\een
The  masses of all of these are determined by   three general $3 \times 3$ complex matrices $p_{pq}$, $g_{pq}$ and $d_{pq}$. Supersymmetry breaking vanishes when either the gauge symmetry breaking vanishes ($p_{pq}=0$); or when the dotspinor mass term vanishes ($d_{pq}=0$).
 The mechanism is such that it is natural to arrange for the observed electron, muon and tau to be far lighter than any of the other charged leptons predicted by the supersymmetry breaking mechanism in cybersusy.  However the magnitude of the breaking needed to be consistent with known experimental results requires that some of the matrix elements in the matrices $p_{pq}$, $g_{pq}$ and $d_{pq}$ are very large or very small.  
\item
The three neutrino flavours work in exactly the same way as the charged leptons except that the matrices $p_{pq}\ra r_{pq}$, $g_{pq}$ and $d_{pq}$
are all different from the matrices for the charged leptons.

\item
We also started a discussion relating to the baryons, but the details there have not been worked out yet.  There is quite a lot of similarity to the case of the leptons however, and it is a reasonable conjecture that the baryons will also be capable of a reasonable spectrum after supersymmetry breaking.  This spectrum needs to be worked out, which is  relatively straightforward, but it is a long and arduous exercize.
\item
Although it appears at first sight that the cybersusy algebra arises for only one composite operator out of the infinite number that can be used to create, say, an electron, there is good reason to believe that it actually applies for all such operators, as was discussed in \ci{cybersusyII} and \ci{cybersusyIII}. This is important, because it tends to emphasize that {\bf the cybersusy mechanism is, in some sense,  unavoidable}, when one looks at composite operators and the composite particles they correspond to.
\item
The nature of cybersusy makes it {\bf clear that there is  no vacuum energy created when supersymmetry is broken}, because the action for the SSM is not changed by the mechanism, and it is well known that gauge symmetry breaking leaves the vacuum energy zero in the SSM.  The supersymmetry breaking takes place in an effective action generated by cybersusy, and, even in that action,  no vacuum energy is generated, because the supersymmetry breaking is more like explicit breaking than like spontaneous breaking.
\item
Discussion of other particles cannot be made without a discussion of the cohomology of the gauge theory in the SSM.  The conservation of baryon number and lepton number means that baryons and leptons cannot have mass mixing with gauge particles, for the purposes of cybersusy, and so the gauge  particles can be ignored in the first approximation.

\een

The mass spectrum for the leptons was displayed and discussed in \ci{cybersusyI} for the special case where there is one flavour only.  The case with three flavours is more complicated, and that case is worked out in detail in this fourth paper.

In \ci{cybersusyI}, we made a number of claims and explained how they worked, but detailed proofs and demonstrations of the claims were postponed to the three succeeding papers: 

\ben
\item
 In  \ci{cybersusyII}, the nilpotent BRS operator for the massless chiral Wess Zumino model was derived and discussed. 
A set of composite chiral dotted spinor pseudosuperfields were generated by certain expressions, which we called simple generators. For these to transform as superfields, certain constraints needed to be satisfied.
\item
In \ci{cybersusyIII}, it was shown that these constraints   can be satisfied in the {\bf massless } SSM, for certain special cases. All of these solutions involve the special spinor field  $\oy_{J \dot \a}$, which gives rise to the special fundamental  pseudosuperfield ${\hat \f}_{J \dot \a}$.  In addition, it turns out that these solutions have the right quantum numbers to look like {\bf massless versions } of composite chiral dotted spinor superfields (dotspinors) for leptons and baryons. 
\item
It was also shown in \ci{cybersusyIII} that, when the gauge symmetry is broken in the SSM, these dotspinors generate the algebra that was used to create the effective cybersusy algebra.
In fact,  these dotspinor superfields cease to be superfields when the gauge symmetry is broken, because they transform with an inhomogeneous term that arises from the $\oy_{J \dot \a}$ term in their composition.  This is the origin of the cybersusy algebra.
\item
To complete the proofs and demonstrations of the claims  in \ci{cybersusyI}, it remains to actually compute the mass matrices and propagators that were used in  \ci{cybersusyI}. Simple versions of these propagators were used in \ci{cybersusyI} to illustrate the model, and some numerical results for those simple versions were reviewed there.  Those simple versions come from the full versions that are computed in this paper.

\een

\subsection{Effective Fields and BRS algebra: Variations $\d_{L}=\d_{\rm LSS}+ \d_{\rm LGSB}$ }
 
\la{leftleptonalgebra}

 The following algebra reflects the BRS cohomology of the neutrinos as discussed in \ci{cybersusyI}.
 Electrons generate the same algebra except for the replacement $
 r_{ q  p } \Ra p_{ q  p }$. In \ci{cybersusyI} we discussed the electrons. Here we discuss the neutrinos.  Everything is the same except the replacement $
 r_{ q  p } \Ra p_{ q  p }$.  As advertised in
 \ci{cybersusyI}, we will write out the cybersusy algebra in components in this paper, whereas in 
\ci{cybersusyI} we wrote it out using superfields.

Here is the left cybersusy algebra for neutrinos:

\be
\d_{L} =
\int d^4 x \;
\lt \{
C^{\e}  \y_{{L}\e}^{p}
\fr{\d}{\d  A_{L}^{p}}
\ebp
+
\lt (
\oC^{\dot \e} \pa_{\a \dot \e} A_{{L}}^{p}
+
F_{{L}}^{p} C_{\a}
\rt )
\fr{\d}{\d 
  \y_{{L}\a}^{p} }
+
\oC^{\dot \e} \pa_{\a \dot \e} \y_{{L}}^{\a p}
\fr{\d}{\d 
  F_{{L}}^{p}}
\ebp
+
\oC^{\dot \e}  \oy_{{L}\dot \e p }
\fr{\d}{\d 
   \A_{{L} p} }
\ebp
\lt (
C^{\e} \pa_{\e \dot \a} \A_{{L} p }
+
\ov F_{{L}p} \oC_{\dot \e}
\rt )
\fr{\d}{\d 
  \oy_{{L}\dot \a p } }
+
C^{\e} \pa_{\e \dot \a} \oy^{\dot \a}_{{L}p}
\fr{\d}{\d 
  \ov F_{{L} p}}
\rt \}
\ee

\be
+
\int d^4 x \;
\lt \{
\lt (
C^{\e}  W_{{L} \e \dot \a  p }
^{}
+
{r}_{q  p }
A_{L}^{q}
\oC_{\dot \a}
\rt )
\fr{\d}{\d 
  \w^{}_{{L} \dot \a  p } }
\ebp
+
\lt (
\oC^{\dot \e} \pa_{\e \dot \e}
\w^{}_{{L} \dot \a  p }
+
C_{\e} 
\L^{}_{{L} \dot \a  p }
-
{r}_{q  p }
\y_{{L}\e}^{q}
\oC_{\dot \a}
\rt )
\fr{\d}{\d 
  W_{{L}\e \dot \a  p } }
\ebp
+
\lt (
\oC_{\dot \e} \pa^{\e \dot \e}
W^{}_{{L} \e \dot \a  p }
+
{r}_{q  p }
F_{L}^{q}
\oC_{\dot \a}
\rt )
\fr{\d}{\d 
  \L^{}_{{L} \dot \a  p }}
\ebp
+
\lt (
\oC^{\dot \e}  \ov W_{{L}  \a \dot \e }^{ p }
+
{\ov r}^{q  p }
\A_{{L} q }
C_{ \a}
\rt )
\fr{\d}{\d \ov \w^{ p }_{{L} \a }}
\ebp
+
\lt (
C^{\e} \pa_{\e \dot \e}
\ov \w^{ p }_{{L} \a}
+
\oC_{\dot \e} 
\ov \L^{ p }_{{L}   \a}
-
{\ov r}^{q  p }
\oy_{{L}\dot \e q}
C_{\a}
\rt )
\fr{\d}{\d   \ov W^{ p }_{{L} \a \dot \e }}
\ebp
  +
\lt (
C_{\e} \pa^{\e \dot \e}
\ov W^{ p }_{{L}  \dot \e  \a }
+
{\ov r}^{q  p }
\ov F_{{L} q}
C_{\a}
\rt )
\fr{\d}{\d\ov \L^{ p }_{{L}   \a } }
\rt \}
\ee

We will divide this into two parts
\be
\d_{L} =\d_{\rm LSS}+ \d_{\rm LGSB} 
\ee
where $\d_{\rm LSS}$ is the part of $\d_{L}$ that remains if
one sets ${ r}_{q  p }={\ov r}^{q  p }=0$.

\subsection{Effective Fields and BRS algebra:  Variations $\d_{R}=\d_{\rm RSS}+ \d_{\rm RGSB}$ }
\la{fergergegeR}

This algebra is the same as the foregoing except for the replacement $L \ra R$.  

\la{rightleptonalgebra}
\be
\d_{R}=
\int d^4 x \;
\lt \{
C^{\e}  \y_{{R}\e}^{ p }
\fr{\d}{\d  A_{R}^{ p }}
\ebp
+
\lt (
\oC^{\dot \e} \pa_{\a \dot \e} A_{{R}}^{ p }
+
F_{{R}}^{ p } C_{\a}
\rt )
\fr{\d}{\d 
  \y_{{R}\a}^{ p } }
+
\oC^{\dot \e} \pa_{\a \dot \e} \y_{{R}}^{\a  p }
\fr{\d}{\d 
  F_{{R}}^{ p }}
\ebp
+
\oC^{\dot \e}  \oy_{{R}\dot \e { p }  }
\fr{\d}{\d 
   \A_{{R} { p } } }
\ebp
\lt (
C^{\e} \pa_{\e \dot \a} \A_{{R} { p }  }
+
\ov F_{{R}{ p } } \oC_{\dot \e}
\rt )
\fr{\d}{\d 
  \oy_{{R}\dot \a { p }  } }
+
C^{\e} \pa_{\e \dot \a} \oy^{\dot \a}_{{R}{ p } }
\fr{\d}{\d 
  \ov F_{{R} { p } }}
\rt \}
\ee

\be
+
\int d^4 x \;
\lt \{
\lt (
C^{\e}  W_{{R} \e \dot \a p }
^{}
+
{r}_{p {q} }
A_{R}^{{q}}
\oC_{\dot \a}
\rt )
\fr{\d}{\d 
  \w^{}_{{R} \dot \a {p} } }
\ebp
+
\lt (
\oC^{\dot \e} \pa_{\e \dot \e}
\w^{}_{{R} \dot \a {p} }
+
C_{\e} 
\L^{}_{{R} \dot \a { p} }
-
{r}_{p {q} }
\y_{{R}\e}^{{q}}
\oC_{\dot \a}
\rt )
\fr{\d}{\d 
  W_{{R}\e \dot \a {p} } }
\ebp
+
\lt (
\oC_{\dot \e} \pa^{\e \dot \e}
W^{}_{{R} \e \dot \a {p} }
+
{r}_{p {q} }
F_{R}^{{q}}
\oC_{\dot \a}
\rt )
\fr{\d}{\d 
  \L^{}_{{R} \dot \a {p} }}
\rt.
\ee

\be
\lt.
+
\lt (
\oC^{\dot \e}  \ov W_{{R}  \a \dot \e }^{{p} }
+
{\ov r}^{p {q} }
\A_{{R} {q} }
C_{ \a}
\rt )
\fr{\d}{\d \ov \w^{{p} }_{{R} \a }}
\ebp
+
\lt (
C^{\e} \pa_{\e \dot \e}
\ov \w^{{p} }_{{R} \a}
+
\oC_{\dot \e} 
\ov \L^{{p} }_{{R}   \a}
-
{\ov r}^{p {q}   }
\oy_{{R}\dot \e {q}}
C_{\a}
\rt )
\fr{\d}{\d   \ov W^{{p} }_{{R} \a \dot \e }}
\ebp
  +
\lt (
C_{\e} \pa^{\e \dot \e}
\ov W^{{p} }_{{R}  \dot \e  \a }
+
{\ov r}^{p {q}  }
\ov F_{{R} {q}}
C_{\a}
\rt )
\fr{\d}{\d\ov \L^{{ p} }_{{R}   \a } }
\rt \}
\ee

We will divide this into two parts
\be
\d_{R} =\d_{\rm RSS}+ \d_{\rm RGSB} 
\ee
where $\d_{\rm RSS}$ is the part of $\d_{R}$ that remains if
one sets ${ r}_{q  p }={\ov r}^{q  p }=0$.

Here are some more definitions of various suboperators and combinations:
\be
\d_{\rm Total} =\d_{\rm L} + \d_{\rm R} 
=\d_{\rm SS}+\d_{\rm GSB}
\ee
\be
\d_{\rm SS} =\d_{\rm LSS} + \d_{\rm RSS} 
\ee
\be
\d_{\rm GSB} =\d_{\rm LGSB} + \d_{\rm RGSB} 
\ee
\be
\d_{\rm L} =\d_{\rm LSS} + \d_{\rm LGSB}
\ee
\be
\d_{\rm R} =\d_{\rm RSS} + \d_{\rm RGSB}
\ee
It is straightforward to verify that:
\be
\d_{\rm Total}^2= \d_{L}^2= \d_{R}^2 = \d_{\rm SS}^2= \d_{\rm LSS}^2= \d_{\rm RSS}^2 =0
\ee
Note that this algebra arises from the cohomology discussion and its application to the SSM in  \ci{cybersusyI}, but really it stands alone too, since it is nilpotent and the action below is invariant in the way described below.

The renormalization of the fields introduces a subtlety relating to the matrices ${r}_{q  p }$ which appear in the algebra.  This subtlety is discussed in Appendix 
\ref{gergrijergerioj}.

\subsection{Action for leptons including the leptonic dotspinor multiplet, invariant under the combined transformation $\d =\d_{\rm SS} + \d_{\rm GSB}
$}
\la{fergergegeAction}

Here we repeat the action from 
\ci{cybersusyI}. The sum of the following terms is invariant under the BRS operators preceding, except for the dotspinor mass term, as we explained in \ci{cybersusyI}:

\be
{\cal A}_{{\rm  ScalarL}}=
\eb
\int d^4 x \;
\lt (
- A^{ p  }_{{L} }
\D \A^{ }_{{L} p  }
-
 \y^{    \a p}_{{L} }
\pa_{\a \dot \b}
 \oy^{ \dot \b}_{{L} p   }
+
 F^{  p }_{{L} }
 \oF^{ }_{{L} p  }
\rt )
\ee

\[
{\cal A}_{{\rm DotspinorL}}=
\int d^4 x \;
\lt (
 \ov \w^{ \a  p }_{{L}  }
\pa_{\a \dot \b}
\D 
\w^{  \dot \b }_{{L}   
 p }
\rt.
\]
\be 
\lt.
-
 \ov W^{ \a \dot \g  p }_{{L}  }
\pa_{\a \dot \b} 
\pa_{\d \dot \g}
W^{  \d \dot \b}_{{L}   
 p  }
-
\ov \L^{ p  \a}_{{L}  }
\pa_{\a \dot \b}
 \L^{  \dot \b}_{{L}   
 p  }
\rt )
\la{rqewfwfwefweL}
\ee

\[
{\cal A}_{\rm KCL1}=
\]
\[
+
\int d^4 x \;
\lt ( 
{\ov r}^{q  p }
\ov \y^{}_{{L}     \dot \a q}
\D 
\w^{  \dot \a}_{{L}   
  p }
+
{\ov r}^{q  p }
\ov F^{}_{{L} q  }
\pa_{\a \dot \a}
W^{  \a \dot \a}_{{L}   
 p  }
\rt.
\]
\be
\lt.
+ 
 {r}_{q  p }
 \y^{q  }_{{L}    \a}
\D 
\ov \w^{   p  \a}_{{L}   
 }
+
 {r}_{q  p }
 F^{ q }_{{L} }
\pa_{\a \dot \a} 
\ov W^{  p  \a \dot \a}_{{L}   
 }
\rt )
\la{wefwefwfwefwL}
 \ee

\[
{\cal A}_{\rm KCL2}=
\]\be
+
\int d^4 x \;
 {r}_{q  p }
{\ov r}^{s  p }
\lt (
  A^{  q  }_{{L} }
\D 
\A^{ }_{{L}   s }
+
 F^{ q   }_{{L} }
 \oF^{ }_{{L}  s  }
\rt )
\ee

\be
{\cal A}_{{\rm ScalarR}}=
\eb
\int d^4 x \;
\lt (
- A^{ { p }   }_{{R} }
\D \A^{ }_{{R} { p }   }
-
 \y^{    \a { p } }_{{R} }
\pa_{\a \dot \b}
 \oy^{ \dot \b}_{{R} { p }    }
+
 F^{  { p }  }_{{R} }
 \oF^{ }_{{R} { p }   }
\rt )
\la{rqewfwfwefweL}
\ee

\[
{\cal A}_{\rm Dotspinor{R}}=
\int d^4 x \;
\lt (
 \ov \w^{ \a {{p}}}_{{R}  }
\pa_{\a \dot \b}
\D 
\w^{  \dot \b }_{{R}   
{{p}}}
\rt. \]
\be 
\lt.
\ebp
-
 \ov W^{ \a \dot \g {{p}}}_{{R}  }
\pa_{\a \dot \b} 
\pa_{\d \dot \g}
W^{  \d \dot \b}_{{R}   
{{p}} }
-
\ov \L^{{{p}} \a}_{{R}  }
\pa_{\a \dot \b}
 \L^{  \dot \b}_{{R}   
{{p}} }
\rt )
\la{rqewfwfwefweR}
\ee

\[
{\cal A}_{{\rm KC}{R}1}=
\]
\[
\int d^4 x \;
\lt ( 
{\ov r}^{ {{p}}q}
\ov \y^{}_{{R}     \dot \a {q}}
\D 
\w^{  \dot \a}_{{R}   
 {{p}}}
+
{\ov r}^{ {{p}}{q}}
\ov F^{}_{{R} {q}  }
\pa_{\a \dot \a}
W^{  \a \dot \a}_{{R}   
{{p}} }
\rt.
\]
\be
\lt.
+ 
 {r}_{ {{p}}{q}}
 \y^{{q}  }_{{R}    \a}
\D 
\ov \w^{  {{p}} \a}_{{R}   
 }
+
 {r}_{ {{p}}{q}}
 F^{ {q} }_{{R} }
\pa_{\a \dot \a} 
\ov W^{ {{p}} \a \dot \a}_{{R}   
 }
\rt )
\la{wefwefwfwefwR}
\ee

\be
{\cal A}_{{\rm KC}{R}2}=
\int d^4 x \;
 {r}_{ {{p}}{q}}
{\ov r}^{{{p}}{s } }
\lt (
  A^{  {q}  }_{{R} }
\D 
\A^{ }_{{R}   {s } }
+
 F^{ {q}   }_{{R} }
 \oF^{ }_{{R}  {s }  }
\rt )
\ee

\be
{\cal A}_{\rm Scalar\; Mass }=
\int d^4 x \;
\eb
 m g_{pq}
\lt (
- \y^{ \a  p}_{{L} }
 \y^{q}_{{R} \a}
+
 A^{  p }_{{L} }
 F^{q}_{{R} }
+
 F^{ p }_{{L} }
 A^{q}_{{R}}
\rt )
\eb+
\int d^4 x \;
\eb
 m \ov g^{p q}
\lt (
- \oy^{\dot \a  }_{{L}p }
 \oy^{}_{{R} \dot \a q}
+
 \A^{   }_{{L} p}
 \ov F^{}_{{R} q}
+
 \ov F^{  }_{{L} p}
 \A^{}_{{R} q}
\rt )
\ee

\be
{\cal A}_{\rm Dotspinor\; Mass }=
\int d^4 x \;
\eb
  m^2 d^{ p  q}  
\lt (
- \w^{\dot \a  }_{{L}  p }
 \L^{}_{{R}\dot \a q}
-
 \L^{\dot \a  }_{{L}  p }
 \w^{}_{{R} \dot \a q}
-
 W^{\a \dot \a  }_{{L}  p }
 W^{}_{{R} \a \dot \a q}
\rt )
\eb
+
\int d^4 x \;
\eb
    m^2 \ov d_{\dot  p  q}  
\lt (
 - \ov \w^{\a  p  }_{{L} }
 \ov \L^{ \ q}_{{R}\a }
-
 \ov \L^{ p  \a  }_{{L}}
 \ov \w^{   q}_{{R}  \a }
-
 \ov W^{  p \a \dot \a  }_{{L} }
 \ov W^{  q}_{{R} \a \dot \a }
\rt )
\ee

\subsection{Invariance of this action}

The operators $\d_{R}$ yield zero of course on the action ${\cal A}_{ {L}}$, and vice versa, simply because they do not involve the same fields.  It is more non-trivial but still straightforward to verify that:
\be
\d_{L} {\cal A}_{{\rm  ScalarL}}
 = 0
\ee
\be
\d_{\rm L} {\cal A}_{{\rm DotspinorL}}=0
\ee
\be
\d_{LSS} 
{\cal A}_{\rm KCL1}+\d_{\rm LGSB}{\cal A}_{{\rm DotspinorL}}=0
\la{wfewfoifrgog1}
\ee
\be
\d_{\rm LSS}
{\cal A}_{\rm KCL2}
+
\d_{\rm LGSB}
{\cal A}_{\rm KCL1}=0
\la{wfewfoifrgog2}
\ee
\be
\d_{\rm LGSB}
{\cal A}_{\rm KCL2}=0
\la{wfewfoifrgog3}
\ee
The Kinetic Compensator terms are put in to achieve invariance when the $\d_{\rm LGSB}$ term is present.  There are similar equations for the right sector.
Then there are the following equations that mix left and right:
\be
\d_{\rm SS} {\cal A}_{\rm Scalar\; Mass }
=0
\ee
\be
\d_{\rm GSB} {\cal A}_{\rm Scalar\; Mass }
=0
\ee
\be
\d_{\rm SS} {\cal A}_{\rm Dotspinor\; Mass }=0
\ee
In fact one can verify that  the complete action is invariant under $\d_{\rm Total}$, except that:
\be
\d_{\rm GSB} {\cal A}_{\rm Dotspinor\; Mass }\neq 0  
\ee
 Can we find compensating terms that will make this invariant, as we did with 
(\ref{wfewfoifrgog1}), (\ref{wfewfoifrgog2}) and (\ref{wfewfoifrgog3})? The answer is {\bf no}.

This is demonstrated in the next subsection.  

\subsection{A demonstration that there is no possible term  
 which can satisfy the requirements for ${\cal A}_{\rm MC1}$ }
\

In this subsection, we shall show that there is no local polynomial solution ${\cal A}_{\rm MC1}$ to the equations:
\be
\d_{\rm GSB} {\cal A}_{\rm Dotspinor\; Mass } + \d_{\rm SS}  {\cal A}_{\rm MC1}
=0\;??
\ee

Consider the spinor mass terms
\[
{\cal A}_{\rm Dotspinor\; Mass }=
\int d^4 x \;
\]
\be  m^2 d^{ p  q}  
\lt (
- \w^{\dot \a  }_{{L}  p }
 \L^{}_{{R}\dot \a q}
-
 \L^{\dot \a  }_{{L}  p }
 \w^{}_{{R} \dot \a q}
-
 W^{\a \dot \a  }_{{L}  p }
 W^{}_{{R} \a \dot \a q}
\rt )
\la{qegregererhfirst}
\ee
\[
+
\int d^4 x \;
\]
\be
    m^2 \ov d_{\dot  p  q}  
\lt (
 - \ov \w^{\a  p  }_{{L} }
 \ov \L^{ \ q}_{{R}\a }
-
 \ov \L^{ p  \a  }_{{L}}
 \ov \w^{   q}_{{R}  \a }
-
 \ov W^{  p \a \dot \a  }_{{L} }
 \ov W^{  q}_{{R} \a \dot \a }
\rt )
\la{qegregererh}
\ee

Here are the relevant variations:
\be
\d_{{\rm GSB \;dotted \;spinor\; part}}
\eb
=
\d_{{\rm LGSB \;dotted \;spinor\; part}}
+
\d_{{\rm RGSB \;dotted \;spinor\; part}}
\ee
where
\be
\d_{{\rm RGSB \;dotted \;spinor\; part}}=
\eb
\int d^4 x \;
\lt \{
\lt (
{r}_{p {q} }
A_{R}^{{q}}
\oC_{\dot \a}
\rt )
\fr{\d}{\d 
  \w^{}_{{R} \dot \a {p} } }
+
\lt (
-
{r}_{p {q} }
\y_{{R}\e}^{{q}}
\oC_{\dot \a}
\rt )
\fr{\d}{\d 
  W_{{R}\e \dot \a {p} } }
\ebp
+
\lt (
{r}_{p {q} }
F_{R}^{{q}}
\oC_{\dot \a}
\rt )
\fr{\d}{\d 
  \L^{}_{{R} \dot \a {p} }}
+
\lt (
{\ov r}^{p {q} }
\A_{{R} {q} }
C_{ \a}
\rt )
\fr{\d}{\d \ov \w^{{p} }_{{R} \a }}
\ebp
+
\lt (
-
{\ov r}^{p {q}   }
\oy_{{R}\dot \e {q}}
C_{\a}
\rt )
\fr{\d}{\d   \ov W^{{p} }_{{R} \a \dot \e }}
  +
\lt (
{\ov r}^{p {q}  }
\ov F_{{R} {q}}
C_{\a}
\rt )
\fr{\d}{\d\ov \L^{{ p} }_{{R}   \a } }
\rt \}
\ee

and

\be
\d_{{\rm LGSB \;dotted \;spinor\; part}}=
\eb
\int d^4 x \;
\lt \{
\lt (
{r}_{q  p }
A_{L}^{q}
\oC_{\dot \a}
\rt )
\fr{\d}{\d 
  \w^{}_{{L} \dot \a  p } }
+
\lt (
-
{r}_{q  p }
\y_{{L}\e}^{q}
\oC_{\dot \a}
\rt )
\fr{\d}{\d 
  W_{{L}\e \dot \a  p } }
\ebp
+
\lt (
{r}_{q  p }
F_{L}^{q}
\oC_{\dot \a}
\rt )
\fr{\d}{\d 
  \L^{}_{{L} \dot \a  p }}
+
\lt (
{\ov r}^{q  p }
\A_{{L} q }
C_{ \a}
\rt )
\fr{\d}{\d \ov \w^{ p }_{{L} \a }}
\ebp
+
\lt (
-
{\ov r}^{q  p }
\oy_{{L}\dot \e q}
C_{\a}
\rt )
\fr{\d}{\d   \ov W^{ p }_{{L} \a \dot \e }}
+
\lt (
{\ov r}^{q  p }
\ov F_{{L} q}
C_{\a}
\rt )
\fr{\d}{\d\ov \L^{ p }_{{L}   \a } }
\rt \}
\ee

The variation of the first part, in 
equation (\ref{qegregererhfirst}),
 is
\be
\d_{{\rm GSB \;dotted \;spinor\; part}}
{\cal A}_{\rm Dotspinor\; Mass }
\eb
\int d^4 x \;
  m^2 d^{ p  q}  
\lt (
- \w^{\dot \a  }_{{L}  p }
 \L^{}_{{R}\dot \a q}
-
 \L^{\dot \a  }_{{L}  p }
 \w^{}_{{R} \dot \a q}
-
 W^{\a \dot \a  }_{{L}  p }
 W^{}_{{R} \a \dot \a q}
\rt )
\eb
=
\int d^4 x \;
  m^2 d^{ p  q}  
\lt (
- \d \w^{\dot \a  }_{{L}  p }
 \L^{}_{{R}\dot \a q}
+ \w^{\dot \a  }_{{L}  p }
 \d \L^{}_{{R}\dot \a q}
\ebp
-
 \d \L^{\dot \a  }_{{L}  p }
 \w^{}_{{R} \dot \a q}
+
 \L^{\dot \a  }_{{L}  p }
 \d \w^{}_{{R} \dot \a q}
\ebp
-
\d W^{\a \dot \a  }_{{L}  p }
 W^{}_{{R} \a \dot \a q}
-
 W^{\a \dot \a  }_{{L}  p }
\d W^{}_{{R} \a \dot \a q}
\rt )
\ee

and this is
\[
\int d^4 x \;
  m^2 d^{ p  q}  
\lt (
- {r}_{s  p }
A_{L}^{s}
\oC^{\dot \a}
 \L^{}_{{R}\dot \a q}
+ \w^{\dot \a  }_{{L}  p }
{r}_{q {q} }
F_{R}^{{q}}
\oC_{\dot \a}
\rt.\]\[
-
{r}_{s  p }
F_{L}^{s}
\oC^{\dot \a}
 \w^{}_{{R} \dot \a q}
+
 \L^{\dot \a  }_{{L}  p }
{r}_{q {q} }
A_{R}^{{q}}
\oC_{\dot \a}
\]
\be
\lt.+
{r}_{q  p }
\y_{{L}}^{\a q}
\oC^{\dot \a} W^{}_{{R} \a \dot \a q}
+
 W^{\a \dot \a  }_{{L}  p }
{r}_{q {q} }
\y_{{R}\a}^{{q}}
\oC_{\dot \a}\rt )
\la{gqgjnryjhyt}
\ee

The variation of the expression in
equation (\ref{qegregererh}) is just the complex conjugate of the above.

Now we must consider what terms ${\cal A}_{\rm MC1}$ can  be added to the action, such that the supersymmetry variations 
 $\d_{{\rm SS}} {\cal A}_{\rm MC1}$ of those terms will cancel the above terms.

It is easy to see that no such terms
 ${\cal A}_{\rm MC1}$ exist, because none of the terms in
 (\ref{gqgjnryjhyt})
can 
arise from the supersymmetry variation of any conceivable term that is consistent with the following criteria:

\ben
\item The new terms ${\cal A}_{\rm MC1}$  have to be made without derivatives to start with, because the troubling terms in (\ref{gqgjnryjhyt})  do not have derivatives.
\la{qergregerge}
\item  They have to have a factor of $m^2$ preceding them because the troubling terms in (\ref{gqgjnryjhyt}) have $m^2$ preceding them. 
\item  They have to have a left and a right field in them because the troubling variations  in (\ref{gqgjnryjhyt}) are of that kind, and the variations do not mix left and right.
\item
They need to have zero fermion number because the action has zero fermion number
and so do the variations.
\item
They need to be local--no terms like $\fr{1 }{\D}$ can be present.
\item
They need to be Lorentz invariant--all Lorentz indices must be contracted.
\item
They need to be bilinear in the fields.
\item
\la{rqggergergre}
Each term in ${\cal A}_{\rm MC1}$ must be a product of one term from $ ( A_L, \y_{L \a}, F_L)$ times one term from $ ( \w_{R\dot \b}, W_{R\g \dot \b}, \L_{R\dot \b})$ or else one term from  
$ ( A_R, \y_{R \a}, F_R)$ times one term from $ ( \w_{L\dot \b}, W_{L\g \dot \b}, \L_{L\dot \b})$ (or else the complex conjugate of this) 
\item 
The criterion above in Point \ref{rqggergergre},
just by itself,  makes it clear that the construction is impossible, because there is no way to contract the indices here to make a Lorentz invariant unless one uses a derivative $\pa_{\a \dot \b}$.  But this is forbidden by Point \ref{qergregerge} above.
  
\een
That concludes the demonstration.

Some comfort with this result can be achieved by trying to write  down  terms and then looking at them.  This does not take very long.  The factor of $m^2$ excludes practically every possibility, because it means that the added terms have to have mass dimension 2, which is very small. 

However there are some terms that come close in some ways.  For example, consider the following possible terms for inclusion in $ {\cal A}_{\rm MC1}$:
\[
 {\cal A}_{\rm MC1}=
\int d^4 x \;
\]
\be
 m^2 
\lt (
x_{L pq} \y^{ \a  p}_{{L} }
 \ov \w^{ q}_{{R}  \a }
+ {\ov x}_L^{p q}
\oy^{\dot \a  }_{{L}p }
\w^{}_{{R} \dot \a q}
+x_{R}^{\dot   p q}
\oy^{\dot \a}_{{R}   p }
 \w^{ }_{{L} q \dot \a }
+{\ov x}_{R \dot  p  q}
 \y^{\a  p }_{{R} }
  \ov \w^{   q}_{{L}\a }
\rt )
\la{gqergqhhtjtjet}
\ee

This expression (\ref{gqergqhhtjtjet})  satisfies the following criteria:
\ben
\item 
This has the right dimension, and a factor of $m^2$ in front.
\item
It has no derivatives.
\item
It is local--it has no terms like $\fr{1 }{\D}$ 
\item
It is Lorentz invariant--there are no uncontracted indices.
\item
It is bilinear in the fields.
\een

However expression (\ref{gqergqhhtjtjet}) fails to satisfy some of the other criteria because:
\ben
\item 
Expression (\ref{gqergqhhtjtjet})
 has non-zero fermion number.
\item
The supersymmety variation of expression (\ref{gqergqhhtjtjet}) is nothing like
what we need to cancel 
(\ref{gqgjnryjhyt})
\item
It does not take the form of a sum of terms each of which is a product of one term from $ ( A_L, \y_{L \a}, F_L)$ times one term from $ ( \w_{R\dot \b}, W_{R\g \dot \b}, \L_{R\dot \b})$ or else one term from  
$ ( A_R, \y_{R \a}, F_R)$ times one term from $ ( \w_{L\dot \b}, W_{L\g \dot \b}, \L_{L\dot \b})$ (or else the complex conjugate of this)
\een

\subsection{The Total Action after integration of the auxiliaries}

\la{qtgrehhtuyrjkyukuy}
As usual in a chiral supersymmetric theory, it is useful to integrate out the auxiliary fields, in order to find an action whose bosonic kinetic and mass terms can be inverted in a straightforward way.  This results here in:

\be
{\cal A}_{{\rm Total\; after\; Integration\; of \; Auxiliaries}}=
\ee

\be
{\cal A}_{{\rm  ScalarL}}=
\eb
\int d^4 x \;
\lt (
- A^{ p  }_{{L} }
\D \A^{ }_{{L} p  }
-
 \y^{    \a p}_{{L} }
\pa_{\a \dot \b}
 \oy^{ \dot \b}_{{L} p   }
\rt )
\ee

\be
{\cal A}_{{\rm DotspinorL}}=
\eb
\int d^4 x \;
\lt (
 \ov \w^{ \a  p }_{{L}  }
\pa_{\a \dot \b}
\D 
\w^{  \dot \b }_{{L}   
 p }
\ebp
-
 \ov W^{ \a \dot \g  p }_{{L}  }
\pa_{\a \dot \b} 
\pa_{\d \dot \g}
W^{  \d \dot \b}_{{L}   
 p  }
-
\ov \L^{ p  \a}_{{L}  }
\pa_{\a \dot \b}
 \L^{  \dot \b}_{{L}   
 p  }
\rt )
\ee

\[
{\cal A}_{{\rm KC}{L}}=
\]
\be
+
\int d^4 x \;
\lt ( 
{\ov r}^{q  p }
\ov \y^{}_{{L}     \dot \a q}
\D 
\w^{  \dot \a}_{{L}   
  p }
+ 
 {r}_{q  p }
 \y^{q  }_{{L}    \a}
\D 
\ov \w^{   p  \a}_{{L}   
 }
\rt )
\ee

\be
+
\int d^4 x \;
 {r}_{q  p }
{\ov r}^{s  p }
\lt (
  A^{  q  }_{{L} }
\D 
\A^{ }_{{L}   s }
\rt )
\ee

\be
{\cal A}_{{\rm ScalarR}}=
\eb
\int d^4 x \;
\lt (
- A^{ { p }   }_{{R} }
\D \A^{ }_{{R} { p }   }
-
 \y^{    \a { p } }_{{R} }
\pa_{\a \dot \b}
 \oy^{ \dot \b}_{{R} { p }    }
\rt )
\ee

\be
{\cal A}_{\rm Dotspinor{R}}=
\eb
\int d^4 x \;
\lt (
 \ov \w^{ \a {{p}}}_{{R}  }
\pa_{\a \dot \b}
\D 
\w^{  \dot \b }_{{R}   
{{p}}}
\ebp
-
 \ov W^{ \a \dot \g {{p}}}_{{R}  }
\pa_{\a \dot \b} 
\pa_{\d \dot \g}
W^{  \d \dot \b}_{{R}   
{{p}} }
-
\ov \L^{{{p}} \a}_{{R}  }
\pa_{\a \dot \b}
 \L^{  \dot \b}_{{R}   
{{p}} }
\rt )
\ee

\be
{\cal A}_{{\rm KC}{R}}=
\]
\[
\int d^4 x \;
\lt ( 
{\ov r}^{ {{p}}q}
\ov \y^{}_{{R}     \dot \a {q}}
\D 
\w^{  \dot \a}_{{R}   
 {{p}}}
+ 
 {r}_{ {{p}}{q}}
 \y^{{q}  }_{{R}    \a}
\D 
\ov \w^{  {{p}} \a}_{{R}   
 }
\rt )
\ee

\be
+
\int d^4 x \;
 {r}_{ {{p}}{q}}
{\ov r}^{{{p}}{s } }
\lt (
  A^{  {q}  }_{{R} }
\D 
\A^{ }_{{R}   {s } }
\rt )
\ee

\be
{\cal A}_{\rm  Mass }=
\int d^4 x \;
 m g_{pq}
\lt (
- \y^{ \a  p}_{{L} }
 \y^{q}_{{R} \a}
\rt )
\ee

\be
+
\int d^4 x \;
\eb
  m^2 d^{ p  q}  
\lt (
- \w^{\dot \a  }_{{L}  p }
 \L^{}_{{R}\dot \a q}
-
 \L^{\dot \a  }_{{L}  p }
 \w^{}_{{R} \dot \a q}
-
 W^{\a \dot \a  }_{{L}  p }
 W^{}_{{R} \a \dot \a q}
\rt )
\ee

\be
{\cal A}_{\rm   Mass }^*=
\int d^4 x \;
 m \ov g^{p q}
\lt (
- \oy^{\dot \a  }_{{L}p }
 \oy^{}_{{R} \dot \a q}
\rt )
\ee

\be
+
\int d^4 x \; 
\eb
    m^2 \ov d_{\dot  p  q}  
\lt (
 - \ov \w^{\a  p  }_{{L} }
 \ov \L^{ \ q}_{{R}\a }
-
 \ov \L^{ p  \a  }_{{L}}
 \ov \w^{   q}_{{R}  \a }
-
 \ov W^{  p \a \dot \a  }_{{L} }
 \ov W^{  q}_{{R} \a \dot \a }
\rt )
\ee

\be 
{\cal A}_{\rm   From\; Auxiliaries }=
\int d^4 x \;
\lt \{
- \lt (
{\ov r}^{p  p }
\pa_{\a \dot \a}
W^{  \a \dot \a}_{{L}   
 p  }
+
 m \ov g^{p q}
 \A^{}_{{R} q}
\rt )
\ebp
({\cal R}^{-1})_{p}^{\;\;t }
\lt (
 {r}_{t r'}
\pa_{\a \dot \a} 
\ov W^{ r' \a \dot \a}_{L}   
+
 m g_{t \dot v}
 A^{\dot v}_{{R}}
\rt ) \rt \}
\ee

\be 
\lt.
- 
\lt (
 (r^T)_{ \dot t r }
\pa_{\b \dot \b} 
\ov W^{ r \b \dot \b}_{R}   
+
 m (g^T)_{ \dot t r}
 A^{r}_{{L}}
\rt )
({\tilde {\cal R}}^{-1})_{\;\; p }^{\dot t }
\ebp
\lt (
(r^{\dag})^{  p  p}
\pa_{\a \dot \a}
W^{  \a \dot \a}_{{R}   
p }
+
 m ( g^{\dag})^{  p  p}
 \A^{}_{{L} p}
\rt )
\rt \}
\ee
Here we use some new  notation, for example:
\be
({\tilde {\cal R}}^{-1})_{\;\; p }^{\dot t }
\ee
This is the inverse of
\be
({\tilde {\cal R}})_{\;\;q_1}^{ p _1}=
( \d +
 {\tilde R})_{\;\;q_1}^{ p _1}
\ee
These symbols and some other notation that we need are summarized in Appendix \ref{notationappendix}. 

Now that we have the action in this form, it is ready for us to calculate the propagators for the fermions and the bosons.  We will start with the fermions.

\section{ The Fermion Kinetic Mixing matrix, the ansatz for its inverse and the Fermion Propagator} 
\la{qftfggerf}

\subsection{Fermionic equations  and Fermionic  6by6 mixing matrix }

Our next task is to find the mass spectrum that corresponds to the action in subsection \ref{qtgrehhtuyrjkyukuy}.  There are plenty of cross terms in the action, both in the bosonic sector and in the fermionic sector.

The kinetic/mass terms give rise by the usual logic to a propagator, and we take the view that the poles of that propagator, as a function of the momentum, yield the masses. 
One could also contemplate an attempt to diagonalize the kinetic/mass action or the propagator, but that seems difficult to accomplish.  The propagator method used here involves some simple algebra (actually, quite a lot of simple algebra) to invert the bosonic and fermionic matrices ${\cal K}_{\rm Fermi} $ introduced below.

\subsection{The fermionic matrix}
\la{erqrregheheh}

We write the fermionic kinetic/mass  terms in the action in subsection \ref{qtgrehhtuyrjkyukuy}
in the form:

\be
{\ov \F}^T{\cal K}_{\rm Fermi}
   \F
=\ee
where

\be
{\ov \F}^T=
\lt (
\begin{array}{cccccc}
\ov \L_{{R}}^{ p_1 \a_1}
&
\w_{{L}  p _2  }^{\dot \a_2}
&
\y_{{L}}^{p_3  \a_3}
&
\oy_{{R} p _4}^{\dot \a_4}
&
\ov \w_{{R}}^{  p_5 \a_5}
&
\L_{{L}  p _6  }^{\dot \a_6}
\end{array}
\rt )
; \F = \lt (
\begin{array}{c}
{ \L}_{{R} q_1 }^{\dot  \b_1} 
\\
{\ov \w}_{{L} }^{q_2 \b_2}
\\
\oy_{{L}q_3}^{ \dot \b_3}
\\
{\y}_{{R} }^{q_4  \b_4 }
\\
{ \w}_{{R}q_5 }^{\dot  \b_5} 
\\
{\ov \L}_{{L} }^{q_6 \b_6}
\end{array}
\rt )
\la{qgqgrgrwehrt}
\ee

\be
{\cal K}_{\rm Fermi} =
\eb
\lt (
 \begin{array}{cccccc}
\begin{array}{|c|}
\hline
(-)
\\
 \pa_{\a_1 \dot  \b_1} 
\\
\d^{\;\;\;q_1}_{p_1} 
\\
\hline
\end{array}
&
\begin{array}{|c|}
\hline
m^2 \\
(d^{\dag})_{p_1 q_2}  
\\
\ve_{   \a_1   \b_2 }
\\
\hline
\end{array}
&0
&0 
&0
&0 
\\
\begin{array}{|c|}
\hline
m^2 \\
d^{ p _2 q_1} 
\\
\ve_{\dot \a_2 \dot \b_1 } 
\\
\hline
\end{array} 
&\begin{array}{|c|}
\hline
\D \pa_{\dot \a_2 \b_2}
\\
 \d_{\;\;q_2}^{ p _2} 
\\
\hline
\end{array} 
&
\begin{array}{|c|}
\hline
(r^{\dag})^{ p _2 q_3} 
\\
\ve_{  \dot \a_2   \dot \b_3} 
\D
\\
\hline
\end{array} 
&0&0&0\\
0& 
\begin{array}{|c|}
\hline
r_{p_3 q_2} 
\\
\ve_{  \a_3   \b_2} 
\D\\
\hline
\end{array} 

&
\begin{array}{|c|}
\hline
-\pa_{ \a_3 \dot \b_3} 
\\
\d_{p_3}^{\;\;q_3}\\
\hline
\end{array} 
 &\begin{array}{|c|}
\hline
 m\\ g_{p_3 q_4} 
\\
\ve_{     \a_3 \b_4 }\\
\hline
\end{array} 
 &0&0\\
0&0&\begin{array}{|c|}
\hline
 m \\(g^{\dag})^{ p _4 q_3 }
\\
\ve_{  \dot \a_4   \dot \b_3} 
\\
\hline
\end{array} 
&\begin{array}{|c|}
\hline
-\pa_{ \b_4 \dot \a_4} 
\\
\d_{\;\;q_4}^{ p _4}\\
\hline
\end{array} 
&\begin{array}{|c|}
\hline
(r^{\dag})^{{  p _4  q_5}} 
\\
\ve_{  \dot \a_4   \dot \b_5} 
\D\\
\hline
\end{array} 
&0\\
&0
&0 
&\begin{array}{|c|}
\hline
r_{{{ p }_5}' {{q}_4}} 
\\
\ve_{    \a_5     \b_4} 
\D
\\
\hline
\end{array} 
& \begin{array}{|c|}
\hline
\D \pa_{\a_5 \dot  \b_5} 
\\
\d^{\;\;\;{ q_5}}_{{p_5}} 
\\
\hline
\end{array} 
&\begin{array}{|c|}
\hline
m^2\\
 (d^{\dag})_{{p_5} q_6}  
\\
\ve_{   \a_5   \b_6 }
\\
\hline
\end{array} 
\\
&0&0&0
&\begin{array}{|c|}
\hline
 m^2\\ d^{ p _6 { q_5}} 
\\
\ve_{\dot \a_6 \dot \b_5 } 
\\
\hline
\end{array} 
&\begin{array}{|c|}
\hline
(-)
\\
\pa_{\dot \a_6 \b_6} 
\\
\d_{\;\;q_6}^{ p _6} \\
\hline
\end{array}  
\\
\end{array}
\rt )
\la{matrixforfermions}
\ee

Now we want to find the propagator that corresponds to this rather complicated kinetic term.

Note that if we set $r=\ov r=0$, this simplifies considerably and represents kinetic/mass terms for three fermions which could be diagonalized using the conversion from two-component Weyl spinors   to Dirac spinors (or just left as Weyl Spinors).
There is an unusual feature even in that case however--the $\w \D \pa \ov \w$ kinetic term involves three derivatives, instead of the usual one that one expects for fermions. This is closely related to our discussion of the $W$ equation of motion in \ci{cybersusyI}.

\subsection{Ansatz for inverse for the fermionic matrix}

\subsection{Notation}

To perform the completion of the square as illustrated in Appendix 
\ref{inversionappendix},  and to generate the propagator and the masses, we  need to find a matrix ${\cal K}_{\rm Fermi} ^{-1} $ which satisfies the following equation: 
\be
{\ov {\F}}^T {\cal K}_{\rm Fermi}  {\cal K}_{\rm Fermi} ^{-1} {\tilde {\ov {\F}}}
=
{\ov \F}^T  
{\tilde {\ov \F}}
\ee
where:

\be
{\ov \F}^T=
\lt (
\begin{array}{cccccc}
\ov \L_{{R}}^{ p_1 \a_1}
&
\w_{{L}  p _2  }^{\dot \a_2}
&
\y_{{L}}^{p_3  \a_3}
&
\oy_{{R} p _4}^{\dot \a_4}
&
\ov \w_{{R}}^{  p_5 \a_5}
&
\L_{{L}  p _6  }^{\dot \a_6}
\end{array}
\rt )
\ee
and we also need source terms which we will write in the form:
\be
{\tilde {\ov \F}} = \lt (
\begin{array}{c}
{\tilde {\ov \L}}_{{R}s_1 \g_1} 
\\
{\tilde {\ov \w}}_{{L}\dot \g_2 }^{ s _2}
\\
{\tilde \y}_{{L} s_3  \g_3}^{}
\\
{\tilde {\ov \y}}_{{R} \dot \g_4 }^{s _4}
\\
{\tilde {\ov \w}}_{{R}s_5   
\g_5} 
\\
{\tilde {\L}}_{{L} \dot \g_6}^{s _6 }
\end{array}
\rt )
\ee

\subsection{Ansatz}
To proceed, we need to make a guess for the form of the propagator.  The following will work:

\be
{\cal K}_{\rm Fermi} ^{-1} =
\fr{1}{m^4}
\eb
\lt (
\begin{array}{cccccc}
\begin{array}{|c|}
\hline
m^2
\\{f_{11}}^{\;\;\;s_1}_{q_1} 
\\
\pa^{\dot \b_1  \g_1   } 
\\
\hline
\end{array}
& 
\begin{array}{|c|}
\hline
m^2 
\\
{f_{12}}_{q_1  s _2} 
\\
\ve^{\dot  \b_1 \dot  \g_2  } 
\\
\hline
\end{array}
&
\begin{array}{|c|}
\hline
m^2 \\
{f_{13}}_{q_1}^{\;\; s_3} 
\\
\;\; \pa^{\dot \b_1 \g_3  } \\
\hline
\end{array}

&
\begin{array}{|c|}
\hline
m^3 \\
{f_{14}}_{q_1  s _4} 
\\
\ve^{ \dot \b_1  \dot \g_4   } 
\\
\hline
\end{array}

&\begin{array}{|c|}
\hline
m \\
{f_{15}}_{q_1}^{\;\;\;s_5} 
\\
\pa^{\dot \b_1 \g_5   } 
\\
\hline
\end{array}
&
\begin{array}{|c|}
\hline
m^3 \\
{f_{16}}_{q_1  s _6} 
\\
\ve^{\dot \b_1 \dot \g_6   } 
\\
\hline
\end{array}

\\
\begin{array}{|c|}
\hline
m^2 \\
 {f_{21}}^{q_2 s_1} 
\\
\ve^{ \b_2  \g_1  } 
\\
\hline
\end{array}
&
\begin{array}{|c|}
\hline
 {f_{22}}^{q_2}_{\;\;\; s _2}
\\
\pa^{\b_2 \dot \g_2   } 
\\
\hline
\end{array}
&
\begin{array}{|c|}
\hline
m^2 \\{f_{23}}^{q_2 s_3}
\\
\ve^{\b_2 \g_3   } 
\\
\hline
\end{array}
&

\begin{array}{|c|}
\hline
m {f_{24}}^{q_2}_{\;\;\; s _4}
\\
\pa^{\b_2 \dot \g_4   } 
\\
\hline
\end{array}

&\begin{array}{|c|}
\hline
m\\ 
{f_{25}}^{q_2  s_5}
\\
\ve^{\b_2 \g_5   } 
\\
\hline
\end{array}

&\begin{array}{|c|}
\hline
m {f_{26}}^{q_2}_{\;\;\;  s _6}
\\
\pa^{\b_2 \dot \g_6   } 
\\
\hline
\end{array}

\\
\begin{array}{|c|}
\hline
m^2 \\{f_{31}}_{q_3}^{ \;\;\;s_1}
\\ 
 \pa^{\dot \b_3 \g_1  } 
\\
\hline
\end{array}

&\begin{array}{|c|}
\hline
m^2 \\ {f_{32}}_{q_3 s _2}
\\ \ve^{\dot \b_3 \dot \g_2 } 
\\
\hline
\end{array}
&
\begin{array}{|c|}
\hline
m^2 \\
{f_{33}}_{q_3}^{ \;\;\;s_3}
\\
\;\;\pa^{\dot \b_3 \g_3 } 
\\
\hline
\end{array}
&\begin{array}{|c|}
\hline
m^3\\
{f_{34}}_{q_3 s _4}
\\
 \ve^{\dot \b_3 \dot \g_4} 
\\
\hline
\end{array}
&

\begin{array}{|c|}
\hline
m \\{f_{35}}_{q_3}^{\;\;\;s_5}

\\ \pa^{ \dot \b_3 \g_5} 
\\
\hline
\end{array}

&
\begin{array}{|c|}
\hline
m^3 \\{f_{36}}_{q_3  s _6}
\\ \ve^{ \dot \b_3 \dot \g_6} 
\\
\hline
\end{array}

 \\

\\
\begin{array}{|c|}
\hline
m^3 \\{f_{41}}^{ q_4 s_1}
\\
\ve^{ \b_4  \g_1 } 
\\
\hline
\end{array}

&

\begin{array}{|c|}
\hline
m \\{f_{42}}^{ q_4}_{\;\;  s _2}
\\
\pa^{ \b_4  \dot \g_2 } 
\\
\hline
\end{array}

&\begin{array}{|c|}
\hline
m^3\\
{f_{43}}^{ q_4 s_3}
\\
 \ve^{\b_4 \g_3} 
\\
\hline
\end{array}&
\begin{array}{|c|}
\hline
m^2\\
{f_{44}}^{ q_4}_{\;\;\; s _4}\\
 \pa^{\b_4 \dot \g_4} 
\\
\hline
\end{array}
&\begin{array}{|c|}
\hline
m^2 \\
{f_{45}^{ q_4  s_5}}\\
\ve^{ \b_4  \g_5 } 
\\
\hline
\end{array}
&\begin{array}{|c|}
\hline
m^2 \\
{f_{46}}^{ q_4}_{\;\;  s _6}\\
\pa^{ \b_4  \dot \g_6 } 
\\
\hline
\end{array}
 \\
\begin{array}{|c|}
\hline
m {f_{51}}_{q_5}^{\;\; \;s_1}
\\ 
\pa^{\dot  \b_5  \g_1 } 
\\
\hline
\end{array}
&

\begin{array}{|c|}
\hline
m \\{f_{52}}_{q_5 s _2}  
\\ 
\ve^{\dot  \b_5  \dot \g_2 } 
\\\hline
\end{array}

&

\begin{array}{|c|}
\hline
m {f_{53}}_{q_5}^{ \;\;\;s_3}
\\ \pa^{\dot \b_5 \g_3 } 
\\
\hline
\end{array}

&
\begin{array}{|c|}
\hline
m^2 \\{f_{54}}_{q_5 s _4} 
\\
\ve^{\dot \b_5 \dot \g_4   } 
\\
\hline
\end{array}&
\begin{array}{|c|}
\hline
  {f_{55}}_{q_5}^{\;\;\;s_5}
\\
\pa^{\dot \b_5 \g_5 } 
\\
\hline
\end{array}
&
\begin{array}{|c|}
\hline
m^2 \\ {f_{56}}_{q_5 s _6} 
\\
\ve^{\dot \b_5 \dot \g_6 } 
\\
\hline
\end{array}
\\
\begin{array}{|c|}
\hline
m^3 \\{f_{61}}^{ q_6  s_1}
\\ 
\ve^{\b_6 \g_1 } 
\\
\hline
\end{array}
&
\begin{array}{|c|}
\hline
m \\{f_{62}}^{ q_6 }_{\;\;\; s _2}
\\ 
\pa^{\b_6 \dot \g_2 } 
\\
\hline
\end{array}
&\begin{array}{|c|}
\hline
m^3 \\{f_{63}}^{ q_6 s_3}
\\ \ve^{\b_6  \g_3 } 
\\
\hline
\end{array}

&
\begin{array}{|c|}
\hline
m^2 \\{f_{64}}^{ q_6}_{\;\;\; s _4}
\\
\pa^{\b_6 \dot \g_4   } 
\\
\hline
\end{array}

& 
\begin{array}{|c|}
\hline
m^2 \\{f_{65}}^{ q_6  s_5}
\\
\ve^{\b_6  \g_5 } 
\\
\hline
\end{array}
&
\begin{array}{|c|}
\hline 
m^2 \\ {f_{66}}^{ q_6 }_{\;\; \;s _6}
\\
\pa^{\b_6 \dot \g_6   } 
\\
\hline
\end{array}
\\
\end{array}
\rt )
\la{fermiansatz}
\ee

 \subsection{ The Fermionic Inverse Matrix  $f_{ij}$ }

By writing out the product of the matrix ${\cal K}_{\rm Fermi} $ with the ansatz for ${\cal K}_{\rm Fermi} ^{-1}$, and setting the result to the unit matrix, one generates six sets of six equations, one set for each column of the following matrix $f_{ij}$.  These are easy to solve and the result is as follows for the coefficients $f_{ij}$ in the ansatz.  The notation is defined in Appendix  
\ref{notationappendix}.

\normalsize

\be
    f_{ij}=
\eb
\lt (
\begin{array}{cccccc}
\begin{array}{|c|}
\hline
- X {\tilde {\cal D}}
\\
- X^2 {\tilde {\cal D}}
\\ . d^{\dag}. r^{\dag}. 
 {\cal J}.  r. 
\\
d.  {\tilde {\cal D}}
\\
\hline
\end{array}
&
\begin{array}{|c|}
\hline
 X^3 d^{\dag}\\
  {\cal D}  .r^{\dag} .
\\
{\cal J}.r.{\cal D}
\\+ d^{\dag}.{\cal D}
\\
\hline
\end{array}
&
\begin{array}{|c|}
\hline
  - X d^{\dag} . \\
{\cal D}. r^{\dag} .  
\\  {\cal J} 
\\
\hline
\end{array}
&
\begin{array}{|c|}
\hline
  - X d^{\dag} .
\\
 {\cal D}. r^{\dag} .
\\
{\tilde {\cal Y}}. 
\\
g .  
  {\tilde {\cal J}} 

\\
\hline
\end{array}
&
\begin{array}{|c|}
\hline
- X^2 d^{\dag} . {\cal D}\\
.r^{\dag} .{\cal J}
\\
.g. {\cal Y} .
\\
r^{\dag} .
  {\tilde {\cal D}}

\\
\hline
\end{array}
&
\begin{array}{|c|}
\hline
-X^2 d^{\dag} .\\ {{\cal D}}. r^{\dag}
  {\tilde {\cal Y}}.\\g.{\tilde {\cal J}}
 .r^{\dag}.\\{\tilde {\cal D}}.d^{\dag}

\\
\hline
\end{array}
 \\
\begin{array}{|c|}
\hline
 { {\cal D}}.d
\\+ X^3  {\cal D}
 .  r^{\dag}. 
\\
 {\cal J}.  r.    {\cal D} .d
\\
\hline
\end{array}
&
\begin{array}{|c|}
\hline
X^3    {\cal D} 
\\
 .r^{\dag} .{\cal J}.r
\\.{\cal D}+ {\cal D}
\\
\hline
\end{array}
&
\begin{array}{|c|}
\hline
  X^2  {\cal D}
\\
. r^{\dag} .  
  {\cal J} 
\\
\hline
\end{array}
&
\begin{array}{|c|}
\hline
-   X  {\cal D}. r^{\dag} .
\\
{\tilde {\cal Y}}.
\\
 g .  
  {\tilde {\cal J}} 

\\
\hline
\end{array}
&
\begin{array}{|c|}
\hline
X^3 {{\cal D}}.r^{\dag} \\
.{\cal J}
.g. {\cal Y}\\ .r^{\dag} .
  {\tilde {\cal D}}

\\
\hline
\end{array}
&
\begin{array}{|c|}
\hline
- X^2  {{\cal D}}\\. r^{\dag}
  {\tilde {\cal Y}}\\.g.{\tilde {\cal J}}
 .r^{\dag}\\.{\tilde {\cal D}}.d^{\dag}

\\
\hline
\end{array}
 \\
\begin{array}{|c|}
\hline
 - X  
 {\cal J}.  r.\\
    {\cal D} .d
\\
\hline
\end{array}
&
\begin{array}{|c|}
\hline
     X^2  {\cal J}.
\\
r.{\cal D}

\\
\hline
\end{array}
&
\begin{array}{|c|}
\hline
  -
  {\cal J} 
\\
\hline
\end{array}
&
\begin{array}{|c|}
\hline
-{\tilde {\cal Y}}.\\
 g .  
  {\tilde {\cal J}} 

\\
\hline
\end{array}
&
\begin{array}{|c|}
\hline
 - X {\cal J}
.g. \\{\cal Y} .r^{\dag} \\.
  {\tilde {\cal D}}

\\
\hline
\end{array}
&
\begin{array}{|c|}
\hline
- X  
  {\tilde {\cal Y}}.\\g.{\tilde {\cal J}}
 .r^{\dag}.\\{\tilde {\cal D}}.d^{\dag}

\\
\hline
\end{array}
 \\
\begin{array}{|c|}
\hline
- X  {\cal Y}
\\ .  g^{\dag}. 
 {\cal J}. 
\\ r.    {\cal D} .d
\\
\hline
\end{array}
&
\begin{array}{|c|}
\hline
    - X   {\cal Y}.g^{\dag}.\\
 {\cal J}.r.{\cal D}

\\
\hline
\end{array}
&
\begin{array}{|c|}
\hline
   - {\cal Y}.\\
 g^{\dag} .  
  {\cal J} 

\\
\hline
\end{array}
&
\begin{array}{|c|}
\hline
-
  {\tilde {\cal J}} 
\\
\hline
\end{array}
&
\begin{array}{|c|}
\hline
X {\cal Y} .r^{\dag}\\ .
  {\tilde {\cal D}}
 \\- X {\cal Y} .g^{\dag}\\ .{\cal J}
.g. {\cal Y} \\.r^{\dag} .
  {\tilde {\cal D}}

\\
\hline
\end{array}
&
\begin{array}{|c|}
\hline
- X  
 {\tilde {\cal J}}
\\ .r^{\dag}.{\tilde {\cal D}}\\
.d^{\dag}

\\
\hline
\end{array}
 \\
\begin{array}{|c|}
\hline
 -  X^2  {\tilde {\cal D}}
\\.r.{\cal Y}
\\
 .  g^{\dag}. 
 {\cal J}.
\\
  r.    {\cal D} .d

\\
\hline
\end{array}
&
\begin{array}{|c|}
\hline
   X^3 {\tilde{\cal D}}
\\
.r.   {\cal Y}.g^{\dag}.
\\
 {\cal J}.r.{\cal D}

\\
\hline
\end{array}
&
\begin{array}{|c|}
\hline
 - X  {\tilde {\cal D}}. 
\\
r.{\cal Y}.
\\
 g^{\dag} .  
  {\cal J} 

\\
\hline
\end{array}
&
\begin{array}{|c|}
\hline
X^2 {\tilde {\cal D}}\\
. r .  
  {\tilde {\cal J}} 

\\
\hline
\end{array}
&
\begin{array}{|c|}
\hline
X^2  {\tilde {\cal D}}.r.{\cal Y} .r^{\dag} .\\
  {\tilde {\cal D}}
\\ - X^2 {\tilde {\cal D}}.r.{\cal Y} .g^{\dag} \\.{\cal J}
.g. {\cal Y}\\
 .r^{\dag} .
  {\tilde {\cal D}}
\\+ {\tilde {\cal D}}

\\
\hline
\end{array}
&
\begin{array}{|c|}
\hline
+
X^3 {\tilde {\cal D}}.\\r. 
 {\tilde {\cal J}}
 .\\r^{\dag}.{\tilde {\cal D}}.d^{\dag}
\\+{\tilde {\cal D}}.d^{\dag}

\\
\hline
\end{array}
 \\
\begin{array}{|c|}
\hline
- X^2 d. {\tilde {\cal D}}
\\
.r.{\cal Y}
 .  g^{\dag}. 
\\
 {\cal J}.  r.  
\\  {\cal D} .d

\\
\hline
\end{array}
&
\begin{array}{|c|}
\hline
 - X^2  d. {\widetilde {\cal D}}\\
.r.  
 {\cal Y}.g^{\dag}\\
. {\cal J}.r.{\cal D}

\\
\hline
\end{array}
&
\begin{array}{|c|}
\hline
  - X d. {\tilde {\cal D}}. 
\\
r.{\cal Y}. g^{\dag} .  
\\
  {\cal J} 

\\
\hline
\end{array}
&
\begin{array}{|c|}
\hline
-X d. {\tilde {\cal D}}
\\
. r .  
  {\tilde {\cal J}} 

\\
\hline
\end{array}
&
\begin{array}{|c|}
\hline
X^2 d. {\tilde {\cal D}}.r.{\cal Y} .r^{\dag}\\ .
  {\tilde {\cal D}}
\\ - X^2 d. {\tilde {\cal D}}.r.{\cal Y} .g^{\dag}\\ .{\cal J}
.g. {\cal Y}\\ .r^{\dag} .
  {\tilde {\cal D}}
\\
+d. {\tilde {\cal D}}

\\
\hline
\end{array}
&
\begin{array}{|c|}
\hline
-
X^2 d. {\tilde {\cal D}}.r. 
 \\{\tilde {\cal J}}
 .r^{\dag}.\\{\tilde {\cal D}}.d^{\dag}
\\- X{\tilde {\cal D}} 

\\
\hline
\end{array}
 \\
\end{array}
\rt )
\la{fmatrixforfermions1}
\ee

\Large

\subsection{Hermiticity constraints on the matrix $f_{ij}$}

This matrix needs to satisfy certain conditions relating to the hermiticity of the matrix ${\cal K}_{\rm Fermi} ^{-1}$.  These are discussed in Appendix \ref{eqrhrtjhynfgrrdfzs}, and they form a useful check on the algebra.

\subsection{Comments on the Fermionic Matrix $f_{ij}$}

In order to find the Fermionic masses, we need to find the values of the variable $X = \fr{\D}{m^2}$ at which any of the
matrix elements of  $f_{ij}$ develop poles as a function
 of $X$.

   The observable masses of the fermions are at the poles  of the matrix elements in $f_{ij}$  above in equation (\ref{fmatrixforfermions1}). Put another way, the value of $X$  where any of the  matrix elements $f_{ij}$  goes to infinity as a function of the variable $X$ is the value of a mass in the theory. $X$ is an abbreviation for:
\be
X = \fr{\D}{m^2} =  \fr{- \pa^{\m}\pa_{\m} }{m^2} 
= \fr{ \pa_{0}\pa_{0}-\pa_{i}\pa_{i} }{m^2}
 \ra    \fr{- p_0^2 + p_i p_i }{m^2}   
\ee

It is not easy to solve this problem using the general form of the matrix (\ref{fmatrixforfermions1}).  However if we take the special case where all the matrices are assumed to commute, and assume that they are all simultaneously diagonalizable, we can make some immediate progress.  It turns out that {\bf every} term in the matrix has the same denominator in that case. 

The demonstration of this fact is relegated to Appendix 
\ref{egrqhrthjtjety}, where it is shown that the fermion propagator for the commuting case has the form: 
\be
{\cal K}_{\rm Fermi} ^{-1} =
\fr{U}{P_{\rm Fermi}}
 {\cal K}_{\rm Fermi\;Numerators} ^{-1}
\ee
where
\be
\fr{U}{P_{\rm Fermi}}
= -\fr{U}{ 
 X 
 \lt \{ X^2 (  U-R) -  D \rt \}^2 
+      G  \lt \{ X^2   U -  D \rt \}^2 }
\la{fermiprop}
\ee
and the expression ${\cal K}_{\rm Fermi\;Numerators} ^{-1}$ has no poles as a function of $X$.

So all the fermion masses are inside the term $\fr{U}{P_{\rm Fermi}}$.  They can be found by finding the zeros of the quintic polynomial equation
\be
P_{\rm Fermi}=  X 
 \lt \{ X^2 (  U-R) -  D \rt \}^2 
+      G  \lt \{ X^2   U -  D \rt \}^2 =0
\ee
 The matrices  $R,G,D,U$  are defined in Appendix 
\ref{notationappendix}.  U is the unit matrix.  $R,G,D$ are all positive definite of course, because they are of the form
\be
R = r. r^{\dag}  , G = g. g^{\dag}  
, D = d. d^{\dag}  
\ee

We shall discuss the masses below after we have calculated similar quantities for the bosons.

\section{ The Boson Kinetic Mixing matrix, the ansatz for its inverse and the Boson Propagator} 

\la{fqwfewhgregher}

\subsection{The Bose Kinetic Term}

\la{erqrreghehehbose}

Define:

\be
{\ov {\cal B}}^T=
\lt (
\begin{array}{llllll}
 \A^{}_{{R}  p _1}
&
W^{\a_2 \dot \a_2  }_{{L}   p _2  }
&
\ov  W^{\a_3 \dot \a_3  p_3}_{{R}  }
&
 A^{ p_4}_{{L} }
\\
\end{array}
\rt )
; 
{\cal B}=\lt (
\begin{array}{lll}
 A^{q_1}_{{R} }
\\
\ov  W^{\b_2 \dot \b_2  q_2}_{{L}  }
\\
W^{\b_3 \dot \b_3  }_{{R}  q_3  }
\\
\ov A^{}_{{L}  q_4}
\\
\end{array}
\rt )
\ee

Then the kinetic/mass action in subsection \ref{qtgrehhtuyrjkyukuy}
 can be written in the form:
\be
{\ov {\cal B}}^T.{\cal K}.
{\cal B}
\ee

where
\be
{\cal K}=
\lt (
\begin{array}{cccc}
\begin{array}{|c|}
\hline
- \D
\\
( \d -
 {\tilde R})_{\;\;q_1}^{ p _1}
\\
- 
 m^2 
\\
(g^{\dag} .   {\cal R}^{-1}
  . g)_{\;\;q_1}^{ p _1}
\\
\hline
\end{array}
& 
\begin{array}{|c|}
\hline
-m \pa_{\b_2 \dot \b_2}
\\
(g^{\dag}.{\cal R}^{-1}.r)^{ p _1}_{\;\; q_2} 
\\
\hline
\end{array}
&0
&0
\\
\\
\begin{array}{|c|}
\hline
+ m \pa_{\a_2 \dot \a_2}
\\
(r^{\dag}.{\cal R}^{-1}.g)^{ p _2}_{\;\; q_1} 
\\
\hline
\end{array}
&
\begin{array}{|c|}
\hline
-\d^{ p _2}_{\;\;q_2}
\\
 \pa_{\a_2 \dot \b_2} 
\pa_{\b_2 \dot \a_2} +
\\
(r^{\dag}.{\cal R}^{-1}.r)^{ p _2}_{\;\;q_2}
\\
 \pa_{\a_2 \dot \a_2} 
\pa_{\b_2 \dot \b_2} 
\\
\hline
\end{array}
&\begin{array}{|c|}
\hline
- m^2 \\
d^{ p _2 q_3} 
\\
\ve_{\dot \a_2 \dot \b_3 } 
\\
\ve_{\a_2 \b_3 } 
\\
\hline
\end{array}
&0
 \\

\\
0
&\begin{array}{|c|}
\hline
- m^2 \\
(d^{\dag})_{  p_3 q_2} 
\\
\ve_{\dot \a_3 \dot \b_2 } 
\\
\ve_{\a_3 \b_2 } 
\\
\hline
\end{array}&
\begin{array}{|c|}
\hline
-\d_{p_3}^{\;\;q_3}
\\
 \pa_{\a_3 \dot \b_3} 
\pa_{\b_3 \dot \a_3}+ 
\\
 (r.{\tilde {\cal R}}^{-1}.r^{\dag})_{p_3}^{\;\;q_3}
\\
 \pa_{\a_3 \dot \a_3} 
\pa_{\b_3 \dot \b_3} 
\\
\hline
\end{array}
&\begin{array}{|c|}
\hline
(r.{\tilde {\cal R}}^{-1}
.g^{\dag})_{p_3}^{\;\;  q_4 } 
\\
m \pa_{ \a_3  \dot \a_3 } 
\\
\hline
\end{array}
 \\
\\
0
&0
& 
\begin{array}{|c|}
\hline
-(g.{\tilde {\cal R}}^{-1}
.r^{\dag})_{p_4}^{\;\;  q_3 } 
\\
m \pa_{ \b_3  \dot \b_3 } 
\\
\hline
\end{array}
&
\begin{array}{|c|}
\hline
- 
 \D
\\
( \d 
-
 R)^{\;\ q_4}_{  p_4}
\\
- 
 m^2 
\\
(g .
{\tilde {\cal R}}^{-1} 
  . g^{\dag})^{\;\ q_4}_{  p_4}
\\
\hline
\end{array}
\\
\end{array}
\rt )
\la{matrixforbosons}
\ee

\subsection{Ansatz for inverse for the bosonic matrix}

To perform the completion of the square as illustrated in Appendix 
\ref{inversionappendix},  and to generate the propagator and the masses, we  need to find a matrix ${\cal K}^{-1} $ which satisfies the following equation: 
\be
{\ov {\cal B}}^T {\cal K} {\cal K}^{-1} {\tilde {\ov {\cal B}}}
=
{\ov {\cal B}}^T  
{\tilde {\ov {\cal B}}}
\ee
where:

\be
{\ov {\cal B}}^T=
\lt (
\begin{array}{llllll}
 \A^{}_{{R}  p _1}
&
W^{\a_2 \dot \a_2  }_{{L}   p _2  }
&
\ov  W^{\a_3 \dot \a_3   p_3}_{{R}  }
&
 A^{ p_4}_{{L} }
\\
\end{array}
\rt )
; {\tilde {\ov {\cal B}}}
=\lt (
\begin{array}{lll}
 {\tilde {\ov A}}^{s _1}_{{R} }
\\
{\tilde { W}}^{{L}  
s _2}_{\g_2 \dot \g_2  }
\\
{\tilde {\ov  W}}^{  }_{{R}\g_3 
\dot \g_3 s_3 }
 \\
{\tilde {A}}^{}_{{L} s_4}
\\
\end{array}
\rt )
\ee

where we will use the ansatz:

\be
{\cal K}^{-1} =
\fr{1}{m^4}
\eb
\lt (
\begin{array}{cccccc}
\begin{array}{|c|}
\hline
m^2 
\\
{b_{11}}^{q_1}_{\;\;s _1} 
\\
\hline
\end{array}
& 
\begin{array}{|c|}
\hline
m \\
{b_{12}}^{q_1}_{\;\; s _2} 
\\
\pa^{\g_2 \dot \g_2 } \\
\hline
\end{array}

&
\begin{array}{|c|}
\hline
m {b_{13}}^{q_1 s_3} 
\\
\pa^{\g_3 \dot \g_3   } 
\\
\\
\hline
\end{array}
&
\begin{array}{|c|}
\hline
m^2 \\
 {b_{14}}^{q_1 s_4} 
\\
\hline
\end{array}

\\
\begin{array}{|c|}
\hline
m \\{b_{21}}^{q_2}_{\;\;\;s _1}
\\ \pa^{\b_2 \dot \b_2 } 
\\
\hline
\end{array}

&
\begin{array}{|c|}
\hline
({b_{221}}^{q_2}_{\;\;s _2}
\\
 \pa^{\b_2 \dot \g_2} 
\pa^{\g_2 \dot \b_2} 
\\
+{b_{222}}^{q_2}_{\;\;s _2}
\\
 \pa^{\g_2 \dot \g_2} 
\pa^{\b_2 \dot \b_2} 
\\
+
 {b_{223}}^{q_2}_{\;\;s _2} m^2
\\
\ve^{\b_2 \g_2}
\ve^{\dot \b_2 \dot \g_2} )
\\
\hline
\end{array}
&\begin{array}{|c|}
\hline
\\
({b_{231}}^{q_2 s_3}
\\
 \pa^{\g_3 \dot \b_2} 
\pa^{\b_2 \dot \g_3} 
\\+{b_{232}}^{q_2 s_3}
\\
 \pa^{\b_2 \dot \b_2} 
\pa^{\g_3 \dot \g_3} 
\\
+{b_{233}}^{q_2 s_3} m^2
\\
\ve^{\dot \b_2 \dot \g_3 } 
\ve^{\b_2 \g_3 } )
\\
\hline
\end{array}
&

\begin{array}{|c|}
\hline
m \\{b_{24}}^{q_2 s_4}
\\ \pa^{\b_2 \dot \b_2 } 
\\
\hline
\end{array}

 \\

\\
\begin{array}{|c|}
\hline
m \\{b_{31}}_{ q_3  s _1}
\\
\pa^{ \b_3  \dot \b_3 } 
\\
\hline
\end{array}

&

\begin{array}{|c|}
\hline
({b_{321}}_{ q_3 s _2}
\\
 \pa^{\b_3 \dot \g_2} 
\pa^{\g_2 \dot \b_3} 
\\
+{b_{322}}_{ q_3 s _2}
\\
 \pa^{\g_2 \dot \g_2} 
\pa^{\b_3 \dot \b_3} 
\\
+ {b_{323}}_{ q_3 s _2} m^2
\\
\ve^{\dot \b_3 \dot \g_2 } 
\ve^{\b_3 \g_2 } )
\\
\hline
\end{array}&
\begin{array}{|c|}
\hline
({b_{331}}_{q_3 }^{\;\;\;s_3}
\\
 \pa^{\b_3 \dot \g_3} 
\pa^{\g_3 \dot \b_3} 
\\
+
{b_{332}}_{q_3}^{\;\;\;s_3}
\\
 \pa^{\g_3 \dot \g_3} 
\pa^{\b_3 \dot \b_3} 
\\
+
m^2 {b_{333}}_{q_3 }^{\;\;\;s_3}
\\
\ve^{\b_3 \g_3}
\ve^{\dot \b_3 \dot \g_3} )
\\
\hline
\end{array}
&\begin{array}{|c|}
\hline
m^2 \\{b_{34}}_{ q_3 }^{\;\;\;s_4}
\\
\pa^{ \b_3  \dot \b_3 } 
\\
\hline
\end{array}
 \\
\\
\begin{array}{|c|}
\hline
m^2 \\{b_{41}}_{ q_4 s _1}
\\ 
\\
\hline
\end{array}
&\begin{array}{|c|}
\hline
m \\{b_{42}}_{ q_4  s _2}
\\ \pa^{\g_2 \dot \g_2 } 
\\
\hline
\end{array}

&
\begin{array}{|c|}
\hline
m \\{b_{43}}_{ q_4 }^{\;\;s_3}
\\
\pa^{\g_3 \dot \g_3   } 
\\
\hline
\end{array}
&
\begin{array}{|c|}
\hline 
m^2 \\ {b_{44}}^{\;\ s_4}_{  q_4}
\\
\hline
\end{array}
\\
\end{array}
\rt )
\la{boseansatz}
\ee

Next, in equation (\ref{bosesoltable}), we will write down the matrix of the coefficients $b_{ij}$ in the following form:

\be
b_{ij}=
\lt (
\begin{array}{cccc}
\begin{array}{|c|}
\hline
b_{11} 
\\
\hline
\end{array}
& 
\begin{array}{|c|}
\hline
b_{12}
\\
\hline
\end{array}
& 
\begin{array}{|c|}
\hline
b_{13}
\\
\hline
\end{array}
& 
\begin{array}{|c|}
\hline
b_{14}
\\
\hline
\end{array}
\\
\\
\begin{array}{|c|}
\hline
b_{21} 
\\
\hline
\end{array}
& 
\begin{array}{|c|}
\hline
b_{221}
\\
\hline
b_{222}
\\
\hline
b_{223}
\\
\hline
\end{array}
& 
\begin{array}{|c|}
\hline
b_{231}
\\
\hline
b_{232}
\\
\hline
b_{233}
\\
\hline
\end{array}
& 
\begin{array}{|c|}
\hline
b_{24}
\\
\hline
\end{array}
\\
\\
\begin{array}{|c|}
\hline
b_{31} 
\\
\hline
\end{array}
& 
\begin{array}{|c|}
\hline
b_{321}
\\
\hline
b_{322}
\\
\hline
b_{323}
\\
\hline
\end{array}
& 
\begin{array}{|c|}
\hline
b_{331}
\\
\hline
b_{332}
\\
\hline
b_{333}
\\
\hline
\end{array}
& 
\begin{array}{|c|}
\hline
b_{34}
\\
\hline
\end{array}
\\
\\
\begin{array}{|c|}
\hline
b_{41} 
\\
\hline
\end{array}
& 
\begin{array}{|c|}
\hline
b_{42}
\\
\hline
\end{array}
& 
\begin{array}{|c|}
\hline
b_{43}
\\
\hline
\end{array}
& 
\begin{array}{|c|}
\hline
b_{44}
\\
\hline
\end{array}
\\
\end{array}
\rt )
\la{bosetemplate}
\ee

The product of ${\cal K} {\cal K}^{-1}$ for the bosons generates four sets of equations, one set for each column of the matrix $b_{ij}$.  The two middle columns have 8 equations and the two outer columns have 4 equations each.  Again the equations are easy to solve and the result is below.  The notation is again defined in Appendix \ref{notationappendix}.

\subsection{The matrix ${b}_{ij} $ }

\be
\la{bosesoltable}
{b}_{ij}  =
\lt (
\begin{array}{cccc}
\begin{array}{|c|}
\hline
- {\widetilde {\cal G}} . 
\\  + 2 X^2 {\widetilde {\cal G}} \\
.  
g^{\dag}.{\cal R}^{-1}.r .
\\ {\cal P}  . r^{\dag}\\.{\cal R}^{-1}.g  \\.{\widetilde {\cal G}}
\\
\hline
\end{array}
& 
\begin{array}{|c|}
\hline
X {\widetilde {\cal G}}.
\\  g^{\dag}.{\cal R}^{-1}.r.
{\cal P}
\\
\hline
\end{array}
& 
\begin{array}{|c|}
\hline
-X {\tilde {\cal G}}.g^{\dag}\\
.{\cal R}^{-1}.r. \\ 
 \fr{U }{\widetilde {\cal V}}
d .{\tilde {\cal P}}
\\
\hline
\end{array}
& 
\begin{array}{|c|}
\hline
-2 X^2  
{\widetilde {\cal G}}
\\
g^{\dag}.{\cal R}^{-1}.r .
\\
  \fr{U}{{\widetilde {\cal V}}} .
d.  \\
 {\tilde {\cal P}} . r.{\tilde {\cal R}}^{-1}
\\
.g^{\dag} .  {\cal G}
\\
\hline
\end{array}
\\
\\
\begin{array}{|c|}
\hline
- X {\cal P}  . r^{\dag}.
\\
{\cal R}^{-1}.g  .{\widetilde {\cal G}}
\\
\hline
\end{array}
& 
\begin{array}{|c|}
\hline
- {\cal D}
\\
\hline
\fr{1}{2}(
{\cal D} 
-  {\cal P}) 
\\
\hline
0
\\
\hline
\end{array}
& 
\begin{array}{|c|}
\hline
0
\\
\hline
  \fr{1}{2 }
 \fr{U }{{\widetilde {\cal V}}}
d .{\tilde {\cal P}}
\\
-   \fr{1}{2 X}
d. {\tilde {\cal D}}
\\
\hline
d. {\tilde {\cal D}}
\\
\hline
\end{array}
& 
\begin{array}{|c|}
\hline
 X \fr{U}{{\widetilde {\cal V}}} .
\\
d.   {\tilde {\cal P}} . 
\\
r.{\tilde {\cal R}}^{-1}
\\
.g^{\dag} .  {\cal G}
\\
\hline
\end{array}
\\
\\
\begin{array}{|c|}
\hline
 X \fr{U}{{\cal V}}
. d^{\dag} 
 \\  . {\cal P}  . r^{\dag}.
\\
{\cal R}^{-1}.g  .{\widetilde {\cal G}}\\
\hline
\end{array}
& 
\begin{array}{|c|}
\hline
0\\
\hline

 \fr{1}{2}\fr{U}{{ {\cal V}}} .
d^{\dag}.
{\cal P}
\\
-  
 \fr{1}{2 X}  
. d^{\dag} . {\cal D}
\\
\hline
- d^{\dag}. {{\cal D}}
\\
\hline
\end{array}
& 
\begin{array}{|c|}
\hline
 -{\widetilde {\cal D}} 
\\
\hline
\fr{1}{2}(
 {\widetilde {\cal D}} 
- {\widetilde {\cal P}}) 
\\
\hline
0
\\
\hline
\end{array}
& 
\begin{array}{|c|}
\hline
- X {\tilde {\cal P}} . 
\\
r.{\tilde {\cal R}}^{-1}
\\
.g^{\dag} .  {\cal G}
\\
\hline
\end{array}
\\
\\
\begin{array}{|c|}
\hline
- 2 X^2 
{\cal G}
 .
\\  g.{\tilde {\cal R}}^{-1}
.r^{\dag} 
  . \fr{U}{{\cal V}}
.\\
 d^{\dag} 
   . {\cal P}  . r^{\dag}.{\cal R}^{-1}.g \\
 .{\widetilde {\cal G}}
\\
\hline
\end{array}
& 
\begin{array}{|c|}
\hline
- X {\cal G}.
g.{\tilde {\cal R}}^{-1}
\\
.r^{\dag}
.
 \fr{U}{{ {\cal V}}} .
\\
d^{\dag}.
{\cal P}
\\
\hline
\end{array}
& 
\begin{array}{|c|}
\hline
X {\cal G}.g
\\
.{\tilde {\cal R}}^{-1}
.r^{\dag} 
\\.
 {\tilde {\cal P}} 
\\
\hline
\end{array}
& 
\begin{array}{|c|}
\hline
-
{\cal G}
\\
+
2 X^2 {\cal G}
\\. g.{\tilde {\cal R}}^{-1}
.r^{\dag}. 
\\  {\tilde {\cal P}} . r.{\tilde {\cal R}}^{-1}
\\.g^{\dag} .  {\cal G}
\\
\hline
\end{array}
\\
\end{array}
\rt )
\la{fwefwefwefwef}
\ee

The hermiticity constraints for this matrix are discussed in Appendix  \ref{qehrthrtjet}.  Again, as in the fermionic case, these form a useful check on the algebra.

\subsection{Discussion of the Commuting Case for the Bosonic Propagator for the Leptons}

For the bosons we also want to find the masses as the poles of the expression for $b_{ij}$.  The situation is similar to that for the fermions in some ways. This is discussed in detail in Appendix \ref{qehrthrtjet}.
The solutions simplify a great deal if one assumes that all the matrices commute.  In that case all the poles can be characterized as either the negative real solutions $X$ of the quartic equation:
\be
X^2 \left(X (1-R)^2+G\right)^2-D \left(G+\left(1-R^2\right) X\right)^2=0
\ee
or  the negative real  solutions $X$ of the quadratic equation
\be
X^2  -D = (X- \sqrt{D} ) (X+ \sqrt{D} ) =0
\ee

This latter expression occurs in the positions $b_{221}$ and $b_{331}$ by itself.  This corresponds to the fact that the mass of the spin one part of the leptonic vector boson is not affected by the mixing at all, and also shows that the mass of the vector boson does not depend on being able to diagonalize the masses. 

As was discussed in \ci{cybersusyI}, the equation
\be
(X- \sqrt{D} )=0
\la{fsfsdafasdfuykykyu}
\ee
does not give rise to a mass.

\subsection{Exact Value of Solutions}
The quartic equation is:
\be
X^2 \left(X (1-R)^2+G\right)^2-D \left(G+\left(1-R^2\right) X\right)^2=0
\ee
and it can be written
\be
X^2 \left(X (1-R)^2+G\right)^2=D \left(G+\left(1-R^2\right) X\right)^2
\ee
and taking the square root of both sides yields two quadratic equations:
\be
X  \left(X (1-R)^2+G\right) =  + \sqrt{D} \left(G+\left(1-R^2\right) X\right)
\ee
\be
X  \left(X (1-R)^2+G\right) =  -\sqrt{D} \left(G+\left(1-R^2\right) X\right)
\ee
and so there are four solutions--two for each of the quadratic equation.  We can write them as follows.  First define:
\be
A_{+} = \sqrt{ B_+^2  + 4 \sqrt{D} G (1 - R)^2}
\ee
\be
B_+ = G +  \sqrt{D}   (-1 + R^2)
\ee
\be
A_{-} = \sqrt{ B_-^2  - 4 \sqrt{D} G (1 - R)^2}
\ee
\be
B_- = G -  \sqrt{D}   (-1 + R^2)
\ee
and then the solutions are given by:
\be
X_1 =  - \fr{A_+ + B_+}{ 2 (1 - R)^2} <0
\ee
\be
X_2 =  \fr{A_+ - B_+}{ 2 (1 - R)^2} >0
\la{fwfwewef4545}
\ee
\be
X_3 =  \fr{-A_- - B_-}{ 2 (1 - R)^2} <0
\ee
\be
X_4 =  \fr{A_- - B_-}{ 2 (1 - R)^2} <0
\ee
where we also indicate their signs, valid for all non-zero positive values of the parameters $G,D$ and $0<R<1$.

The three negative solutions correspond to boson masses.  The positive solution (\ref{fwfwewef4545}) 
 does not give rise to a mass--it is like the term $(X-\sqrt{D})$ that was discussed in \ci{cybersusyI}, and the discussion there applies to (\ref{fwfwewef4545}) also.

So we can write
\be
X^2 \left(X (1-R)^2+G\right)^2-D \left(G+\left(1-R^2\right) X\right)^2\eb
= (1-R^2)^2(X- X_1)(X- X_2)(X- X_3)(X- X_4)
\ee
and we could rewrite the entire matrix 
(\ref{fwefwefwefwef}) in terms of partial fractions that look more like conventional propagators, except that there are also the terms like (\ref{fwfwewef4545})
and (\ref{fsfsdafasdfuykykyu}), which have the wrong signs to give rise to masses.

The same is true for the fermions of course, but it may well be that there is no closed algebraic form for the solutions for that quintic polynomial.

\subsection{Terms like $b_{11}$: Are there other poles in the Matrix (\ref{bosesoltable})?}

For the commuting case, there are no other poles for negative real $X$ in the matrix 
(\ref{bosesoltable}).  This is similar to the situation for 
subsection \ref{ergerhrthrth}
 which came up while we considered the fermi matrix. In other words, no new pole is present for the boson matrix,  because in such cases the new pole that might be present cancels against a similar term in the numerator, as in subsection \ref{ergerhrthrth}.
This is discussed further in Appendix \ref{qehrthrtjet}.

\section{Conclusion}

\subsection{Masses for the Leptonic Bosons and Fermions after Supersymmetry Breaking  }

\la{geioupeuirgeu}

\subsubsection{One Flavour}

As stated in \ci{cybersusyI}, we have found that for one flavour, the fermion masses are at the poles of the propagator
 $\fr{U}{P_{\rm Fermi}}$.  They can be found by finding the negative real solutions for the quintic polynomial equation:
\be
P_{\rm Fermi}=  X 
 \lt \{ X^2 (  U-R) -  D \rt \}^2 
+      G  \lt \{ X^2   U -  D \rt \}^2 =0
\ee
It appears, from the numerical results in \ci{cybersusyI},  that there are three negative real solutions for this equation, at least for a wide range of the parameters.

Also, as stated in \ci{cybersusyI}, we have found that for one flavour, the boson masses are at the poles of the Bose propagators. They can be found by finding the three negative real solutions for the quartic polynomial equation:
\be
P_{\rm Bose}= X^2  
\lt (  X  (U-R)^2  +  G  
\rt )^2 - \lt ( X(U -R^2)+G \rt )^2  D 
=0
\la{efgjihnengben}
\ee
and at the negative real solution for the quadratic polynomial equation:
\be
P_{\rm D}=X^2 U - D=0 
\la{fdaehtrhrthbose}\ee

\subsubsection{Three Flavours and comparison with experiment}

To actually make a comparison with experimental results, and it is possible there may be some reasonably soon at the LHC, one would need to deal with the general forms of the propagators.  That looks quite difficult, unless one assumes that the matrices commute, in which case the propagators reduce to three copies of the one-flavour results.

For the neutrinos, there evidently is something to say about dark matter candidates here.  For neutrinos and electrons, it appears that something needs to be said about the early universe too.  

In \ci{cybersusyI}, we presented three possible choices of the numbers for one flavour, and noted that it seemed to be possible to avoid any conflict with current experimental knowledge. This statement may need to be revised in view of things like the renormalization group etc., which we have not tried to cover here. It also needs to be considered whether there are other issues that raise difficulties for this method of supersymmetry breaking. In \ci{cybersusyIII}, it was conjectured, and partially shown, that {\bf all} relevant operators for the leptons and the baryons, not just the ones we looked at, will obey the cybersusy algebra.  If that is not so, then that would be a  big problem for the mechanism.

\appendix

\section{Inversion of the Kinetic/Mass Actions}
\la{inversionappendix}

The fermions and the bosons in this paper both have 
 complicated mixed up kinetic/mass actions, and we write them in  matrix form above in subsection \ref{erqrregheheh}
 for the fermions and subsection \ref{erqrreghehehbose} for the bosons. 

This matrix form is as follows for both cases:

\be
\int d^4 x \; {\cal L}
\ee
where 
\be
 {\cal L} = {\ov \F}^T{\cal K}_{\rm }   \F
\ee
and the generating functional has the form
\be 
{\cal Z} = \int \d \F \d {\ov \F}  e^{i \int d^4 x \lt \{
 {\ov \F}^T{\cal K}_{\rm }   \F+ {\ov \F}^T { \tilde {\ov \F}}+ { \F}^T {\tilde \F} \rt \}}
\ee

Suppose that we can find an operator 
${\cal K}_{\rm }  ^{-1}$ such that:
\be
{\ov \F}^T{\cal K}_{\rm }  
 {\cal K}_{\rm }  ^{-1} {\tilde \F}
=
{\ov \F}^T U {\tilde \F}
=
{\ov \F}^T {\tilde \F}
\ee
where $U=U^{\dag}$ is  a kind of unit operator.
Then we perform the transformation
\be
\F \ra \F + 
{\cal K}_{\rm }  ^{-1} {\tilde {\ov \F}}
\ee
\be
{\ov \F}^T \ra {\ov \F}^T + 
 { {\tilde \F}}^T {\cal K}_{\rm }  ^{-1 \dag}
\ee
The generating functional then has the form:
\be 
{\cal Z} = \int \d \F \d {\ov \F} 
e^{i \int d^4 x \lt \{
 \lt (  {\ov \F}^T + 
 { {\tilde \F}}^T {\cal K}_{\rm }  ^{-1 \dag } \rt )
{\cal K}_{\rm }  \lt (
\F + 
{\cal K}_{\rm }  ^{-1} {\ov {\tilde \F}}
\rt )
-
 { {\tilde \F}}^T {\cal K}_{\rm }  ^{-1 \dag } 
{\cal K}_{\rm }  
{\cal K}_{\rm }  ^{-1} {\ov {\tilde \F}}
  \rt \}}
\ee
Hermiticity implies that
\be
{\cal K}_{\rm }   = -{\cal K}_{\rm }  ^{\dag}
\ee
\be
({\cal K}_{\rm }  ^{-1})^{\dag} = -{\cal K}_{\rm }  ^{-1}
\ee
and so we have
\be
{\cal K}_{\rm }  
 {\cal K}_{\rm }  ^{-1}
= U
\ee
and 
\be
 {\cal K}_{\rm }  ^{-1 \dag}
{\cal K}_{\rm }  
= -U^{\dag} = -U
\ee
and 
so after the integration $\int \d \F \d {\ov \F}$, which just supplies an overall factor, we are left with 

\be 
{\cal Z} =
e^{-i \int d^4 x  \lt \{
 { {\tilde \F}}^T {\cal K}_{\rm }  ^{-1 \dag } 
{\cal K}_{\rm }  
{\cal K}_{\rm }  ^{-1} {\ov {\tilde \F}}
  \rt \}}
\eb
=e^{i \int d^4 x  \lt \{
 { {\tilde \F}}^T {\cal K}_{\rm }  ^{-1 } 
{\cal K}_{\rm }  
{\cal K}_{\rm }  ^{-1} {\ov {\tilde \F}}
  \rt \}}
\eb
= e^{+i \int d^4 x  \lt \{
{ {\tilde \F}}^T{\cal K}_{\rm }  ^{-1 } 
 {\ov {\tilde \F}} \rt \}}
\ee
for the generating functional.  This is the form which contains the propagator ${\cal K}_{\rm }  ^{-1}$, and the matrix $U$ functions as a charge conjugation operator.

\section{Collection of Notation used for the effective action for the leptons}

\la{notationappendix}

\subsection{Compound Matrices}

We use the notation $U$ for the unit matrix, and whatever indices are appropriate for the context.

Here are the definitions of the $R,G,D$ matrices:
\be
{r}_{ {{p}}{s }}
{\ov r}^{{{q}}{s } }
=
{r}_{ {{p}}{s }}
{r^{\dag}}^{ s  q }
= R_{p }^{\;\;q }
\ee

\be
{g}_{ {{p}}{q}}
{\ov g}^{ s q }
={g}_{ {{p}}{q}}
{g^{\dag}}^{ q s }
= {G}_{p }^{\;\; s }
\ee

\be
{d}_{ {{ p }}{q}}
{\ov d}^{{\dot {s}}{    q} }
= {d}_{ {{ p }}{q}} {d}^{\dag{    q} {\dot {s}} }
= {D}_{ p }^{\;\;s  }
\ee
Here are the definitions of the ${\widetilde R},{\widetilde G},{\widetilde D}$ matrices:

\be
{r}_{ {{p}}{q}}
{\ov r}^{{{p}}{s } }
={r^{\dag}}^{ s  p }
{r}_{ {{p}}{q}}
= {\widetilde R}_{\;\;q }^{s  }
\ee

\be
{g}_{ {{p}}{q}}
{\ov g}^{{{p}}{s } }
={g^{\dag}}^{ s  p }
{g}_{ {{p}}{q}}
= {\widetilde G}_{\;\;q }^{s  }
\ee

\be
{d}_{ {{ p }}{q}}
{\ov d}^{{\dot {p}}{    s} }
={d^{\dag}}^{  s p  }
{d}_{ {\dot {p}}{  q}}
= {\widetilde D}_{\;\;q }^{s }
\ee

and we also need the following combinations:

\be
(\tilde {\cal R}_{-})_{\;\;q_1}^{ p _1}=
( \d -
 {\tilde R})_{\;\;q_1}^{ p _1}
\ee
 
\be
({\tilde {\cal R}})_{\;\;q_1}^{ p _1}=
( \d +
 {\tilde R})_{\;\;q_1}^{ p _1}
\ee
 
\be
({\cal R}_{-})_{p_4}^{\;\;q_4}=
( \d -
 R)_{p_4}^{\;\;q_4}
\ee
 
\be
 ({\cal R})_{p_3}^{\;\;q_3}=
( \d +
 R)_{p_3}^{\;\;q_3}
\ee

\subsubsection{Matrices that arise for the Bosons}

\be 
{\cal D} = \fr{U}{ X^2 U - d . d^{\dag}}
\ee
\be 
{\widetilde {\cal D}} = \fr{U}{ X^2 U -   d^{\dag} . d}
\ee

\be
{  {\cal G}} = \fr{U}{( X  {{\cal R}}_{-} 
+  
g .   {\tilde {\cal R}}^{-1}
  . g^{\dag} )}
\ee

\be
{\widetilde {\cal G}} = \fr{U}{( X  {\widetilde {\cal R}}_{-} 
+  
g^{\dag} .   {\cal R}^{-1}
  . g )}
\ee

\be {\cal V} = 
X \lt (
 2 \;  r.{\tilde {\cal R}}^{-1}
.g^{\dag} .
\fr{U}{\lt ( X {\cal R}_{-} 
+ 
g .
{\tilde {\cal R}}^{-1} 
  . g^{\dag} 
\rt )}.
  g.{\tilde {\cal R}}^{-1}
.r^{\dag} 
+
\lt (
 U
- 2 
 r.{\tilde {\cal R}}^{-1}.r^{\dag}
\rt )
\rt )
\ee
and
\be 
{\widetilde {\cal V}}=X \lt \{2 \; r^{\dag}.{\cal R}^{-1}.g . \fr{U}{  \lt (  X  {\tilde R}_{-} 
+ 
g^{\dag} .   {\cal R}^{-1}
  . g \rt ) }.
g^{\dag}.{\cal R}^{-1}.r 
+ 
\lt (
 U
-
2 
r^{\dag}.{\cal R}^{-1}.r 
\rt ) 
\rt \}
\ee

\be
{  {\cal P}}
=
\fr{1}{X} \lt (\fr{U}{\widetilde {\cal V} - d . \fr{U}{{ {\cal V}}} .d^{\dag}  } \rt)
\ee

\be
{\tilde {\cal P}}
=
\fr{1}{X} \fr{U}{{\rm Denominator}}
=
\fr{1}{X} \lt (\fr{U}{ {\cal V} - d^{\dag} .\fr{U}{{\widetilde {\cal V}}} .d  } \rt)
\ee

\be
{\rm Denominator}=
 {\cal V} - d^{\dag} \fr{U}{{\widetilde {\cal V}}} .d  
\ee

\be
{\cal M} = \lt \{
r^{\dag}.{\cal R}^{-1}.g . \fr{U}{  \lt (  X  {\tilde R}_{-} 
+ 
g^{\dag} .   {\cal R}^{-1}
  . g \rt ) }.
g^{\dag}.{\cal R}^{-1}.r 
-
r^{\dag}.{\cal R}^{-1}.r  
\rt \}
\eb
=  \fr{1}{2 X}\lt ({\widetilde {\cal V}}- U X \rt ) 
\ee

\subsubsection{Matrices that arise for the Fermions}

\be
{\cal D}
=
\fr{ U}{
 X^2 -d. d^{\dag}  }
=
\fr{ U}{
 X^2 -D  }
\la{wegiogonbreoib}
\ee

\be
{\tilde {\cal  D}}
=
\fr{ U}{
 X^2 -  d^{\dag}.d  }
=
\fr{ U}{
 X^2 -{\tilde D}  }
\ee

\be
{\cal Y}
=\fr{U}{ \lt \{
U
-  r^{\dag}
\fr{X^2 U}{
 X^2  -
 d^{\dag} d
} . 
  r
\rt \}}
\ee

\be
{\tilde {\cal Y}}
=
\fr{U}{\lt \{
U - 
r. \fr{U X^2}{ (X^2 U - d.  d^{\dag})} 
.  r^{\dag} 
\rt \}}
\ee

\be
{\cal J}=  \fr{U }{ X 
{\widetilde{\cal Y}} + g .
 {\cal Y}^{-1} .g^{\dag} } 
\eb
=
{ \lt [ U X -   r. 
\fr{X^3 U}{X^2 U - d. d^{\dag} }.
  r^{\dag}. 
+ g. \fr{U}{ \lt \{
U
-  r^{\dag}
\fr{X^2 U}{
 X^2  -
 d^{\dag} d
} . 
  r
\rt \}}
 . 
 g^{\dag}
\rt ]}
\la{fqewfwetbjjp}
\ee

\be
{\tilde {\cal  J}}
=  \fr{U }{ X 
{{\cal Y}} + g^{\dag} .
 {\widetilde{\cal Y}}^{-1} .g} 
\eb
=
\fr{U}{\lt ( X  U  
-  r^{\dag} . \fr{X^3 U}{ (X^2  U 
- d^{\dag} . d)} . 
r + g^{\dag} . \fr{U}{\lt \{
U - 
r. \fr{U X^2}{ (X^2 U - d.  d^{\dag})} 
.  r^{\dag} 
\rt \}} . g
\rt )} 
\ee

\section{Regarding the Matrix ${\cal K}^{-1}_{\rm Fermi}$}

\la{eqrhrtjhynfgrrdfzs}

${\cal K}^{-1}_{\rm Fermi}$ is introduced in Chapter \ref{qftfggerf}. In this Appendix we collect some detailed calculations that arise when verifying the calculation of the Matrix ${\cal K}^{-1}_{\rm Fermi}$, and some related matters.

\subsection{Inverses and some Relations}

Consider the inverses of some of the matrices defined in appendix \ref{notationappendix}:

\be
{\cal J}^{-1}= { \lt [ U X +   r. 
\fr{X^3 U}{d. d^{\dag} 
- X^2 }.
  r^{\dag}. 
+ g. \fr{U}{ \lt \{
U
-  r^{\dag}
\fr{X^2 U}{
 X^2  -
 d^{\dag} d
} . 
  r
\rt \}}
 . 
 g^{\dag}
\rt ]}
\ee
and
\be
{\tilde {\cal  J}}^{-1}
=
{\lt ( X  U  
-  r^{\dag} . \fr{X^3 U}{ (X^2  U 
- d^{\dag} . d)} . 
r + g^{\dag} . \fr{U}{\lt \{
U - 
r. \fr{U X^2}{ (X^2 U - d.  d^{\dag})} 
.  r^{\dag} 
\rt \}} . g
\rt )} 
\ee

\be
{\cal Y}^{-1}
= { \lt \{
U
-  r^{\dag}
\fr{X^2 U}{
 X^2  -
 d^{\dag} d
} . 
  r
\rt \}}
\ee
and
\be
{\tilde {\cal Y}}^{-1}
=
{\lt \{
U - 
r. \fr{U X^2}{ (X^2 U - d.  d^{\dag})} 
.  r^{\dag} 
\rt \}}
\ee

Note that
\be
{  {\cal J}}^{-1}= X {\tilde {\cal Y}}^{-1} + g. {{\cal Y}} . g^{\dag}
\ee
and
\be
{\tilde  {\cal J}}^{-1}= X {  {\cal Y}}^{-1} + g^{\dag}. {\tilde {\cal Y}} . g
\la{f6rwerwerw}
\ee

\subsubsection{ A Basic Identity}

We will now prove the identity
\be
{{\cal Y}}. g^{\dag} { {\cal J}}=
\lt ({\tilde {\cal Y}} .g . {\tilde {\cal J}}\rt )^{\dag}
\ee

To prove this, note that:
\be
\lt ({\tilde {\cal Y}} .g  . {\tilde {\cal J}}\rt )^{\dag}=
{\tilde {\cal J}} .g^{\dag}. {\tilde {\cal Y}}
\ee
So the above identity that requires proof is equivalent to
\be
{{\cal Y}}. g^{\dag} { {\cal J}}=
{\tilde {\cal J}} .g^{\dag}. {\tilde {\cal Y}}
\ee
This is equivalent to the inverse of the same identity:
\be
\lt ( {{\cal Y}}. g^{\dag} { {\cal J}}
\rt )^{-1}
=
\lt ({\tilde {\cal J}} .g^{\dag}. {\tilde {\cal Y}}\rt )^{-1}
\la{infwefefwe}
\ee
The left side of (\ref{infwefefwe}) is

\be
\lt ( {{\cal Y}}. g^{\dag} { {\cal J}}
\rt )^{-1}
=
 {{\cal J}}^{-1}. (g^{\dag})^{-1} { {\cal Y}}^{-1}
\eb
=\lt ( X {\tilde {\cal Y}}^{-1} + g. {{\cal Y}} . g^{\dag}\rt ). (g^{\dag})^{-1} { {\cal Y}}^{-1}
\eb
=  X {\tilde {\cal Y}}^{-1}.(g^{\dag})^{-1} { {\cal Y}}^{-1} + g. {{\cal Y}} . g^{\dag} . (g^{\dag})^{-1} { {\cal Y}}^{-1}
\eb
=  X {\tilde {\cal Y}}^{-1}.(g^{\dag})^{-1} { {\cal Y}}^{-1} + g
\ee
and right side of (\ref{infwefefwe}) is
\be
\lt ({\tilde {\cal J}} .g^{\dag}. {\tilde {\cal Y}}\rt )^{-1}
=
{\tilde {\cal Y}}^{-1} .(g^{\dag})^{-1} . {\tilde {\cal J}}^{-1}
\eb
={\tilde {\cal Y}}^{-1} .(g^{\dag})^{-1} \lt (
X {  {\cal Y}}^{-1} + g^{\dag}. {\tilde {\cal Y}} . g \rt )
\eb
={\tilde {\cal Y}}^{-1} .(g^{\dag})^{-1} 
X {  {\cal Y}}^{-1} + {\tilde {\cal Y}}^{-1} .(g^{\dag})^{-1} .g^{\dag}. {\tilde {\cal Y}} . g 
\eb
=X {\tilde {\cal Y}}^{-1} .(g^{\dag})^{-1} 
 {  {\cal Y}}^{-1} +  g 
\ee
The left and right sides of   (\ref{infwefefwe}) are therefore equal. 
So we have proved the identity:
\be
{{\cal Y}}. g^{\dag} { {\cal J}}=
\lt ({\tilde {\cal Y}} .g . {\tilde {\cal J}}\rt )^{\dag}
={\tilde {\cal J}} .g^{\dag}. {\tilde {\cal Y}}
\la{theidentitty}
\ee

\subsection{Diagonal Elements have equivalent matrices in pairs}

These remarks help to familiarize the large matrix 

\subsubsection{Similarity of  $
f_{66}\approx f_{11}$}

Note that the following two matrix elements are closely related.  They give rise to the same masses.
\be 
 {f_{11}}
=   - X   {\tilde {\cal D}} 
- X^2    d^{\dag}
.{{\cal D}}  .
 r^{\dag} . 
  {{\cal J}}  .
 r.  
{{\cal D}} .d 
  \ee

\be 
 {f_{66}}
=  - X {\tilde {\cal D}}  - X^2 
d.{\tilde {\cal D}}.  r   .{\tilde{\cal J}} . r^{\dag} .
 {\tilde {\cal D}}. d^{\dag}
  \ee

\subsubsection{Similarity of $
f_{33}\approx f_{44}$}

Note that the following two matrix elements are closely related.  They give rise to the same masses.

\be
f_{33}=-{\cal J}
\ee
\be
f_{44}=-{\widetilde {\cal J}}
\ee

\subsubsection{Similarity of  $
f_{55}\approx f_{22}$}

The following two matrix elements are also closely related but it is not so obvious as it was for ${f_{66}}$ and ${f_{11}}$:
\be 
 {f_{55}}
=  {\tilde {\cal D}}
  -X^2  {\tilde {\cal D}}.r. {{\cal Y}}. g^{\dag}. {{\cal J}}    .g.
  {{\cal Y}}.r^{\dag}.{\tilde {\cal D}}
+X^2 {\tilde {\cal D}}.r.{{\cal Y}}.r^{\dag}.{\tilde {\cal D}}
 \ee

\be 
 {f_{22}}
= X^3   {{\cal D}}.  r^{\dag} .
{{\cal J}}.
 r. {{\cal D}}
 +   {{\cal D}}
 \ee

To show that they are closely related, we 
use \ref{theidentitty}:

\be
{{\cal Y}}. g^{\dag} { {\cal J}}=
\lt ({\tilde {\cal Y}} .g . {\tilde {\cal J}}\rt )^{\dag}
={\tilde {\cal J}} .g^{\dag}. {\tilde {\cal Y}}
\ee

Using these we get:
\be 
 {f_{55}}
=  {\tilde {\cal D}}
  -X^2  {\tilde {\cal D}}.r.
( {{\cal Y}}. g^{\dag}. {{\cal J}})  
  .g.
  {{\cal Y}}.r^{\dag}.{\tilde {\cal D}}
+X^2 {\tilde {\cal D}}.r.{{\cal Y}}.r^{\dag}.{\tilde {\cal D}}
\ee
\be 
=  {\tilde {\cal D}}
  -X^2  {\tilde {\cal D}}.r.
({\tilde {\cal J}} .g^{\dag}. {\tilde {\cal Y}})  
  .g.
  {{\cal Y}}.r^{\dag}.{\tilde {\cal D}}
+X^2 {\tilde {\cal D}}.r.{{\cal Y}}.r^{\dag}.{\tilde {\cal D}}
\ee
\be 
=  {\tilde {\cal D}}
  -X^2  {\tilde {\cal D}}.r.
{\tilde {\cal J}} .
(g^{\dag}. {\tilde {\cal Y}}    .g)
.
  {{\cal Y}}.r^{\dag}.{\tilde {\cal D}}
+X^2 {\tilde {\cal D}}.r.{{\cal Y}}.r^{\dag}.{\tilde {\cal D}}
\ee
Now use
(\ref{f6rwerwerw}):
\be
{\tilde  {\cal J}}^{-1}= X {  {\cal Y}}^{-1} + g^{\dag}. {\tilde {\cal Y}} . g
\ee
which implies
\be
{\tilde  {\cal J}}^{-1}- X {  {\cal Y}}^{-1} = g^{\dag}. {\tilde {\cal Y}} . g
\ee
to get 
\be 
 {f_{55}}
= {\tilde {\cal D}}
  -X^2  {\tilde {\cal D}}.r.
{\tilde {\cal J}} .
({\tilde  {\cal J}}^{-1}- X {  {\cal Y}}^{-1})
.
  {{\cal Y}}.r^{\dag}.{\tilde {\cal D}}
+X^2 {\tilde {\cal D}}.r.{{\cal Y}}.r^{\dag}.{\tilde {\cal D}}
\ee
which is
\be 
 {f_{55}}
= {\tilde {\cal D}}
  -X^2  {\tilde {\cal D}}.r.
{\tilde {\cal J}} .
({\tilde  {\cal J}}^{-1})
.
  {{\cal Y}}.r^{\dag}.{\tilde {\cal D}}
\eb
  -X^2  {\tilde {\cal D}}.r.
{\tilde {\cal J}} .
(- X {  {\cal Y}}^{-1})
.
  {{\cal Y}}.r^{\dag}.{\tilde {\cal D}}
+X^2 {\tilde {\cal D}}.r.{{\cal Y}}.r^{\dag}.{\tilde {\cal D}}
\ee
which is then
\be
= {\tilde {\cal D}}
  -X^2  {\tilde {\cal D}}.r.
.
  {{\cal Y}}.r^{\dag}.{\tilde {\cal D}}
  +X^3  {\tilde {\cal D}}.r.
{\tilde {\cal J}} .
 .r^{\dag}.{\tilde {\cal D}}
+X^2 {\tilde {\cal D}}.r.{{\cal Y}}.r^{\dag}.{\tilde {\cal D}}
\ee
or
\be
{f_{55}}= {\tilde {\cal D}}
  +X^3  {\tilde {\cal D}}.r.
{\tilde {\cal J}} 
 .r^{\dag}.{\tilde {\cal D}}
\ee
to compare with
\be 
 {f_{22}}
= X^3   {{\cal D}}.  r^{\dag} .
{{\cal J}}.
 r. {{\cal D}}
 +   {{\cal D}}
 \ee
These are related by  tilde operations and clearly have the same eigenvalues.

\subsection{Detailed Verification that the matrix $f_{ij}$ yields a hermitian ${\cal K}$}

We shall go through the calculations in some detail here in a few places. 

\ben
\item
{Proof that $f_{14} = (f_{41})^{\dag}$  }

\be f_{41}=    - X   {\cal Y}.g^{\dag}.
 {\cal J}.r.{\cal D}.d
\ee
\be
f_{14}=  - X d^{\dag} .
 {\cal D}. r^{\dag} .
{\tilde {\cal Y}}. 
g .  
  {\tilde {\cal J}} 
\ee
So
\be f_{41}^{\dag}=       - X d^{\dag} .
 {\cal D}. r^{\dag}. {\cal J}.g.
 {\cal Y} 
\ee
and using \ref{theidentitty}
\be
{{\cal Y}}. g^{\dag} { {\cal J}}=
\lt ({\tilde {\cal Y}} .g . {\tilde {\cal J}}\rt )^{\dag}
={\tilde {\cal J}} .g^{\dag}. {\tilde {\cal Y}}
\ee
which implies
\be
{{\cal J}}. g . { {\cal Y}}
={\tilde {\cal Y}} .g. {\tilde {\cal J}}
\ee
we get
\be f_{41}^{\dag}=       - X d^{\dag} .
 {\cal D}. r^{\dag}. {\tilde {\cal Y}} .g. {\tilde {\cal J}}
= f_{14} 
\ee
so the hermiticity is proved .

\item{Proof that $f_{15} = (f_{51})^{\dag}$ }

 \be
f_{15}=
X^2 d^{\dag} . {\cal D}\\
.r^{\dag} .{\cal J}
\\
.g. {\cal Y} .
\\
r^{\dag} .
  {\tilde {\cal D}}
\ee

 \be
f_{51}=
 -  X^2  {\tilde {\cal D}}
\\.r.{\cal Y}
\\
 .  g^{\dag}. 
 {\cal J}.
\\
  r.    {\cal D} .d
\ee

This is obvious.

\item{Proof that $f_{16} = (f_{61})^{\dag}$ }

\be
f_{16} =   
-X^2 d^{\dag} .\\ {{\cal D}}. r^{\dag}
  {\tilde {\cal Y}}.\\g.{\tilde {\cal J}}
 .r^{\dag}.\\{\tilde {\cal D}}.d^{\dag}
\ee
\be
f_{61} =  
 - X^2  d. { {\cal D}}\\
.r.  
 {\cal Y}.g^{\dag}\\
. {\cal J}.r.{\cal D}
\ee
This is obvious using
\be
{{\cal J}}. g . { {\cal Y}}
={\tilde {\cal Y}} .g. {\tilde {\cal J}}
\ee

\item{Proof that $f_{24} = (f_{42})^{\dag}$  }

This is obvious using the identity \be
{{\cal J}}. g . { {\cal Y}}
={\tilde {\cal Y}} .g. {\tilde {\cal J}}
\ee

\item{Proof that $f_{25} = (f_{52})^{\dag}$ }

This is obvious 
just from looking at the matrix elements.

\item{Proof that $f_{26} = (f_{62})^{\dag}$}

\be
f_{26} =   
- X^2  {{\cal D}}\\. r^{\dag}
  {\tilde {\cal Y}}\\.g.{\tilde {\cal J}}
 .r^{\dag}\\.{\tilde {\cal D}}.d^{\dag}
\ee
\be
f_{62} =  
 - X^2  d. { {\cal D}}\\
.r.  
 {\cal Y}.g^{\dag}\\
. {\cal J}.r.{\cal D}
\ee
This is obvious using the identity \be
{{\cal J}}. g . { {\cal Y}}
={\tilde {\cal Y}} .g. {\tilde {\cal J}}
\ee

\item{Proof that $f_{34} = (f_{43})^{\dag}$ }

\be
f_{43} =  -{{\cal Y}}. g^{\dag} { {\cal J}}
\ee
and
\be
f_{34} = -{\tilde {\cal Y}} .g . {\tilde {\cal J}}
\ee
satisfy the identity
\be
f_{43} = (f_{34})^{\dag}
\ee

This follows immediately from (\ref{theidentitty}).

\item{Proof that $f_{36} = (f_{63})^{\dag}$ }

\be 
f_{63}
=
  - X d. {\tilde {\cal D}}. 
\\
r.{\cal Y}. g^{\dag} .  
\\
  {\cal J} 
\ee
\be 
f_{36}
=
- X  
  {\tilde {\cal Y}}.\\g.{\tilde {\cal J}}
 .r^{\dag}.\\{\tilde {\cal D}}.d^{\dag}
\ee

This is obvious using the identity \be
{{\cal J}}. g . { {\cal Y}}
={\tilde {\cal Y}} .g. {\tilde {\cal J}}
\ee

\item{Hermiticity for  $
f_{45}= f_{54}^{\dag}$}

We need to show that
\be 
 {f_{45}}
=  
  -X   {{\cal Y}}. g^{\dag}. {{\cal J}}    .g.
  {{\cal Y}}.r^{\dag}.{\tilde {\cal D}}
+X {{\cal Y}}.r^{\dag}.{\tilde {\cal D}}
 \ee
and
\be 
 {f_{54}}
=  X^2  {\tilde {\cal D}}.  r . {\tilde{\cal J}}
  \ee
satisfy
\be 
 {f_{45}}^{\dag}
=
 {f_{54}}
\ee
We note that
\be 
 {f_{45}}^{\dag} 
=  
  -X   {\tilde {\cal D}}.r.{{\cal Y}}.g^{\dag}.{{\cal J}}.g .{{\cal Y}}
+X {\tilde {\cal D}} .r.{{\cal Y}}
 \ee
so to prove hermiticity we need to show that
\be 
-X  {{\cal Y}}.g^{\dag}.{{\cal J}}.g .{{\cal Y}}
+ X {{\cal Y}}
=X^2  {\tilde{\cal J}}
 \ee
Use the identity
\be
{{\cal Y}}. g^{\dag} { {\cal J}}=
{\tilde {\cal J}} .g^{\dag}. {\tilde {\cal Y}}
\ee
to convert this to
\be 
-X  
{\tilde {\cal J}} .g^{\dag}. {\tilde {\cal Y}}.g .{{\cal Y}}
+ X {{\cal Y}}=X^2  {\tilde{\cal J}}
\ee
Now we recall that:
\be
{\tilde  {\cal J}}^{-1}= X {  {\cal Y}}^{-1} + g^{\dag}. {\tilde {\cal Y}} . g
\ee
The latter implies that:
\be
{\tilde  {\cal J}}. {\tilde  {\cal J}}^{-1}.{\cal Y}= 
{\tilde  {\cal J}} \lt (X {  {\cal Y}}^{-1} 
+ g^{\dag}. {\tilde {\cal Y}} . g\rt ).{\cal Y}
\ee
which is
\be
{\cal Y}= 
X {\tilde  {\cal J}}  
+ {\tilde  {\cal J}}   . g^{\dag}. {\tilde {\cal Y}} . g .{\cal Y}
\ee
or 
\be
-  {\tilde  {\cal J}}   . g^{\dag}. {\tilde {\cal Y}} . g .{\cal Y}+ {\cal Y}= 
X {\tilde  {\cal J}}  
\ee
which is just what we need.

\item{Proof that $f_{46} = (f_{64})^{\dag}$ }

\be 
f_{46}
=
- X  
 {\tilde {\cal J}}
\\ .r^{\dag}.{\tilde {\cal D}}\\
.d^{\dag}
\ee
\be 
f_{64}
=
-X d. {\tilde {\cal D}}
\\
. r .  
  {\tilde {\cal J}} 
\ee
is obvious.

\item{Proof that $f_{56} = (f_{65})^{\dag}$ }

\be 
 {f_{65}}
=  d.{\tilde {\cal D}}
  -X^2  d.{\tilde {\cal D}}.r. {{\cal Y}}. g^{\dag}. {{\cal J}}    .g.
  {{\cal Y}}.r^{\dag}.{\tilde {\cal D}}
\eb
+X^2 d.{\tilde {\cal D}}.r.{{\cal Y}}.r^{\dag}.{\tilde {\cal D}}
 \ee

\be 
 {f_{56}}
=  {\tilde {\cal D}}.  d^{\dag} + X^3 
{\tilde {\cal D}}.  r   .{\tilde{\cal J}} . r^{\dag} .
 {\tilde {\cal D}}. d^{\dag}
  \ee

First let us simplify ${f_{65}}$:
Use:
\be
{\tilde  {\cal J}}^{-1}= X {  {\cal Y}}^{-1} + g^{\dag}. {\tilde {\cal Y}} . g
\ee
The latter implies that:
\be
{\tilde  {\cal J}}. {\tilde  {\cal J}}^{-1}.{\cal Y}= 
{\tilde  {\cal J}} \lt (X {  {\cal Y}}^{-1} 
+ g^{\dag}. {\tilde {\cal Y}} . g\rt ).{\cal Y}
\ee
which is
\be
{\cal Y}= 
X {\tilde  {\cal J}}  
+ {\tilde  {\cal J}}   . g^{\dag}. {\tilde {\cal Y}} . g .{\cal Y}
\ee
and use \ref{theidentitty}
\be
{{\cal Y}}. g^{\dag} { {\cal J}}=
\lt ({\tilde {\cal Y}} .g . {\tilde {\cal J}}\rt )^{\dag}
={\tilde {\cal J}} .g^{\dag}. {\tilde {\cal Y}}
\ee
which implies
\be
{{\cal J}}. g . { {\cal Y}}
={\tilde {\cal Y}} .g. {\tilde {\cal J}}
\ee
so we have
\be
{\cal Y}= 
X {\tilde  {\cal J}}  
+ {\tilde  {\cal J}}   . g^{\dag}. {\tilde {\cal Y}} . g .{\cal Y}
=
X {\tilde  {\cal J}}  
+ {{\cal Y}}. g^{\dag} { {\cal J}}. g .{\cal Y}
\ee
which implies that
\be
 {{\cal Y}}. g^{\dag} { {\cal J}}. g .{\cal Y}
={\cal Y}- 
X {\tilde  {\cal J}}  
\ee
Now put this into
\be 
 {f_{65}}
=  d.{\tilde {\cal D}}
  -X^2  d.{\tilde {\cal D}}.r. ({{\cal Y}}. g^{\dag}. {{\cal J}}    .g.
  {{\cal Y}}).r^{\dag}.{\tilde {\cal D}}
\eb
+X^2 d.{\tilde {\cal D}}.r.{{\cal Y}}.r^{\dag}.{\tilde {\cal D}}
\eb
=  d.{\tilde {\cal D}}
  -X^2  d.{\tilde {\cal D}}.r. ( {\cal Y}- 
X {\tilde  {\cal J}}  ).r^{\dag}.{\tilde {\cal D}}
\eb
+X^2 d.{\tilde {\cal D}}.r.{{\cal Y}}.r^{\dag}.{\tilde {\cal D}}
\eb
=  d.{\tilde {\cal D}}
  -X^2  d.{\tilde {\cal D}}.r. (  - 
X {\tilde  {\cal J}}  ).r^{\dag}.{\tilde {\cal D}}
\eb
  -X^2  d.{\tilde {\cal D}}.r. ( {\cal Y}   ).r^{\dag}.{\tilde {\cal D}}
\eb
+X^2 d.{\tilde {\cal D}}.r.{{\cal Y}}.r^{\dag}.{\tilde {\cal D}}
\eb
=  d.{\tilde {\cal D}}
  +X^3  d.{\tilde {\cal D}}.r.   {\tilde  {\cal J}}  .r^{\dag}.{\tilde {\cal D}}
 \ee
and now the hermiticity is obvious.
\een

\subsection{Calculations relating to the absence of certain propagators in the commuting case, even though they look like they might be there}
\la{egrqhrthjtjety}

As mentioned in Chapter \ref{qftfggerf}, in the commuting case, it is possible to show that there is really only one kind of denominator for every term in $f_{ij}$, namely
\be
\fr{U}{P_{\rm Fermi}} = \fr{ U }{  X 
 \lt \{ X^2 (  U-R) -  D \rt \}^2 
+      G  \lt \{ X^2    -  D \rt \}^2   }
\la{qewgyhjuyjk89ou}
\ee
  This seems to require detailed calculation to see.  There ought to be a better way to see it, but the author has not found that better way.  Here are some of the details.

\subsubsection{Simplifications from Commuting Matrices}

\be
{\cal D}
={\tilde {\cal  D}}
=
\fr{ U}{
 X^2 -D }
\ee

\be
{\cal Y}
={\tilde {\cal Y}}
=\fr{U}{ \lt \{
U
-  r^{\dag}
\fr{X^2 U}{
 X^2  -
 d^{\dag} d
} . 
  r
\rt \}}
\Ra
\fr{(X^2 - D)}{
X^2 (U-R)
-D}  
\ee

\be
{\cal J}= 
{\tilde {\cal  J}}
=  
\fr{ \lt \{ X^2 (  U-R) -  D \rt \}
 \lt \{ X^2    -  D \rt \}}{ 
 X 
 \lt \{ X^2 (  U-R) -  D \rt \}^2 
+      G  \lt \{ X^2    -  D \rt \}^2 }
\la{whtjekjyukur}
\ee

\subsubsection{Putting the term $f_{11} $ together for commuting matrices}

\be
f_{11} = -X { {\cal D}}
- X^2 { {\cal D}}^2
 D R 
 {\cal J} 
=
 -X { {\cal D}}
( U
+ X { {\cal D}}
 D R 
 {\cal J} )
\ee
We find for the commuting case that:
\be
f_{11} =
-X 
\lt (   
\fr{ 
 \lt \{ X^2 (  U-R) -  D \rt \} 
  X (  U-R) 
+      G  \lt \{ X^2    -  D \rt \} 
 }{ 
 X 
 \lt \{ X^2 (  U-R) -  D \rt \}^2 
+      G  \lt \{ X^2    -  D \rt \}^2 }
\rt )
\ee
and, as advertised, the factor of $\fr{ U}{
 X^2 -D }$ has been cancelled and does not 
give rise to a pole here.

\subsection{Another example: ${f}_{43} =    {\cal Y} 
 .g^{\dag} .{\cal J}$}
\la{ergerhrthrth}

We have claimed that there are no other poles for negative real $X$ in the matrix 
(\ref{fmatrixforfermions1}).  At first it appears that there might be a new pole corresponding, say,  to the expression ${f}_{43} =    {\cal Y} 
 .g^{\dag} .{\cal J}$.  Let us look at that more closely.  We will again restrict ourselves to the case where all the matrices commute. 

For that case, the expression 
\be
{\cal Y}
=\fr{U}{ \lt \{
U
-  r^{\dag}
\fr{X^2 U}{
 X^2  -
 d^{\dag} d
} . 
  r
\rt \}}
\Ra
\fr{U}{ \lt \{
U
-  R
\fr{X^2 U}{
 X^2  -
 D
} 
\rt \}}
\eb
\Ra
\fr{ X^2  -
 D}{ \lt \{
 X^2 (U-R) - D
\rt \}}
\ee
certainly has a pole by itself at the zero of
\be
 X^2 (U-R) - D
=0.
\ee
  However when multiplied by ${\cal J}$, this pole disappears. 

The value of ${\cal J}$ for the commuting case was set down in 
(\ref{whtjekjyukur}). It was:

\be
- {\cal J}= -\fr{ \lt \{ X^2 (  U-R) -  D \rt \}
 \lt \{ X^2   U -  D \rt \}}{ 
 X 
 \lt \{ X^2 (  U-R) -  D \rt \}^2 
+      G  \lt \{ X^2   U -  D \rt \}^2 }
\ee

So we see that the product 
\be
{\cal Y} {\cal J}= \fr{ X^2  -
 D}{ \lt \{
 X^2 (U-R) - D
\rt \}}\lt (
\fr{ \lt \{ X^2 (  U-R) -  D \rt \}
 \lt \{ X^2   U -  D \rt \}}{ 
 X 
 \lt \{ X^2 (  U-R) -  D \rt \}^2 
+      G  \lt \{ X^2   U -  D \rt \}^2 }
\rt )
\eb
= 
\fr{ \lt \{ X^2 U  - D\rt \}^2
}{ 
 X 
 \lt \{ X^2 (  U-R) -  D \rt \}^2 
+      G  \lt \{ X^2   U -  D \rt \}^2 }
\ee
and, as advertised, there is no pole at the zero of 
\be
 X^2 (U-R) - D
=0.
\ee

Similar considerations apply to all the other matrix elements ${f}_{ij}$.

\subsubsection{Commuting Case for $ {f_{22}}= X^3   {{\cal D}}^2.  R .
{{\cal J}}.
 +   {{\cal D}}
$}

For the commuting case we start with:
\be 
 {f_{22}}
= X^3   {{\cal D}}^2.  R .
{{\cal J}}.
 +   {{\cal D}}
 \ee

and use
\be
{\cal D}
={\tilde {\cal  D}}
=
\fr{ U}{
 X^2 -D }
\ee

\be
{\cal Y}
={\tilde {\cal Y}}
=\fr{U}{ \lt \{
U
-  r^{\dag}
\fr{X^2 U}{
 X^2  -
 d^{\dag} d
} . 
  r
\rt \}}
\Ra
\fr{(X^2 - D)}{
X^2 (U-R)
-D}  
\ee

\be
{\cal J}= 
{\tilde {\cal  J}}
=  
\fr{ \lt \{ X^2 (  U-R) -  D \rt \}
 \lt \{ X^2    -  D \rt \}}{ 
 X 
 \lt \{ X^2 (  U-R) -  D \rt \}^2 
+      G  \lt \{ X^2    -  D \rt \}^2 }\ee

We find that:
\be
{f_{22}}
=
\fr{  
 \lt \{ X^2 (  U-R) -  D \rt \}
+      G  \lt \{ X^2    -  D \rt \} 
  }{  X 
 \lt \{ X^2 (  U-R) -  D \rt \}^2 
+      G  \lt \{ X^2    -  D \rt \}^2   }
 \ee
and once again the denominator is the usual one for this ${f_{22}}$ (and so for ${f_{55}}$ also).

\subsubsection{Various Combinations}

 A careful look at $f_{ij}$ shows that there are a number of terms that have products of the matrices ${\cal D}
{\cal Y}
{\cal J} 
$ in various combinations.  In this  subsubsection we examine several combinations.

Thus 
\be
{\cal J} 
  {\cal D} 
\ee
is
\be
\fr{ U}{
 X^2 -D }
\fr{ \lt \{ X^2 (  U-R) -  D \rt \}
 \lt \{ X^2    -  D \rt \}}{ 
 X 
 \lt \{ X^2 (  U-R) -  D \rt \}^2 
+      G  \lt \{ X^2    -  D \rt \}^2 }\ee
and this is
\be
\fr{ \lt \{ X^2 (  U-R) -  D \rt \}
 }{ 
 X 
 \lt \{ X^2 (  U-R) -  D \rt \}^2 
+      G  \lt \{ X^2    -  D \rt \}^2 }\ee
and we see that the factor of $\fr{ U}{
 X^2 -D }$ is gone.

The product
\be
{\cal D}
{\cal Y}
{\cal J}= 
\ee

\be
=
\fr{ U}{
 X^2 -D }
\ee

\be
\fr{(X^2 - D)}{
X^2 (U-R)
-D}  
\ee
\be
\fr{ \lt \{ X^2 (  U-R) -  D \rt \}
 \lt \{ X^2    -  D \rt \}}{ 
 X 
 \lt \{ X^2 (  U-R) -  D \rt \}^2 
+      G  \lt \{ X^2    -  D \rt \}^2 }\ee
It cancels two problems and again yields the usual denominator:
\be
=
\fr{ 
 \lt \{ X^2    -  D \rt \}}{ 
 X 
 \lt \{ X^2 (  U-R) -  D \rt \}^2 
+      G  \lt \{ X^2    -  D \rt \}^2 }\ee

The product
\be
{\cal Y}
{\cal J}= 
\ee

\be
\fr{(X^2 - D)}{
X^2 (U-R)
-D}  
\ee

\be
\fr{ \lt \{ X^2 (  U-R) -  D \rt \}
 \lt \{ X^2    -  D \rt \}}{ 
 X 
 \lt \{ X^2 (  U-R) -  D \rt \}^2 
+      G  \lt \{ X^2    -  D \rt \}^2 }\ee

\be
=
\fr{  
 \lt \{ X^2    -  D \rt \}^2 }{ 
 X 
 \lt \{ X^2 (  U-R) -  D \rt \}^2 
+      G  \lt \{ X^2    -  D \rt \}^2 }\ee
and once again the denominator is the familiar one.

Using the above we get
\be
{\cal D}^2
{\cal Y}
{\cal J}= 
\ee

\be
=
\lt (\fr{ U}{
 X^2 -D }\rt )^2
\fr{  
 \lt \{ X^2    -  D \rt \}^2 }{ 
 X 
 \lt \{ X^2 (  U-R) -  D \rt \}^2 
+      G  \lt \{ X^2    -  D \rt \}^2 }
\ee
\be
=
\fr{  
U}{ 
 X 
 \lt \{ X^2 (  U-R) -  D \rt \}^2 
+      G  \lt \{ X^2    -  D \rt \}^2 }\ee
and once again it is the familiar denominator.

All the terms in $f_{ij}$ respond in the same manner, with the result advertised in (\ref{qewgyhjuyjk89ou}).

\section{Regarding the Matrix ${\cal K}^{-1}_{\rm Bose}$}

\la{qehrthrtjet}

\subsection{General Remarks on Hermiticity}

For those elements of $b_{ij}$ which get into ${\cal K}^{-1}_{\rm Bose}$ with one derivative, hermiticity requires that
\be
b_{ij} = - b_{ji}^{\dag}
\ee
For those elements of $b_{ij}$ which get into ${\cal K}^{-1}_{\rm Bose}$ with two or no derivatives, hermiticity requires that
\be
b_{ij} = + b_{ji}^{\dag}
\ee

\subsection{Main Diagonal}

Clearly this divides into two sets.

\subsection{Off Diagonal}

\ben

\item{Verify that $b_{12}=- b_{21}^{\dag}$}
This appears in \ref{boseansatz} in the term
\be
m \\
{b_{12}}^{q_1}_{\;\; s _2} 
\\
\pa^{\g_2 \dot \g_2 } \\
\ee
which has one derivative and that gives rise to the 
requirement
$b_{12}=- b_{21}^{\dag}$. 
We have
\be
b_{12}=X {\widetilde {\cal G}}.
\\  g^{\dag}.{\cal R}^{-1}.r.
{\cal P}
\ee
\be
b_{21}=- X {\cal P}  . r^{\dag}.
\\
{\cal R}^{-1}.g  .{\widetilde {\cal G}}
\ee
The condition  $b_{12}=- b_{21}^{\dag}$ is clearly satisfied.

\item{Verify that $b_{13}= - b_{31}^{\dag}$}
This appears in \ref{boseansatz} in the term
\be
m {b_{13}}^{q_1 s_3} 
\\
\pa^{\g_3 \dot \g_3   } 
\ee
which has one derivative and that gives rise to the 
requirement
$b_{13}= -b_{31}^{\dag}$.
We have
\be
b_{13}=-
X {\tilde {\cal G}}.g^{\dag}\\
.{\cal R}^{-1}.r. \\ 
 \fr{U }{\widetilde {\cal V}}
d .{\tilde {\cal P}}
\ee
\be
b_{31}=
 X \fr{U}{{\cal V}}
. d^{\dag} 
 \\  . {\cal P}  . r^{\dag}.
\\
{\cal R}^{-1}.g  .{\widetilde {\cal G}}\\
\ee

We want to show that
\be
{b}_{13}
=
-{b}_{31}^{\dag}
\la{erqghergheurtyt}
\ee

We have
\be
{b}_{31}^{\dag}
=
 X {\widetilde {\cal G}}
. g^{\dag}. {\cal R}^{-1}
 \\.r  . {\cal P}   
.d   .
\fr{U}{{\cal V}}
\ee
so equation (\ref{erqghergheurtyt}) reduces to
\be
\fr{U }{\widetilde {\cal V}}
d .{\tilde {\cal P}}?=?
{\cal P}   
.d   .
\fr{U}{{\cal V}}
\ee
or equivalently the inverse, which is
\be
\lt (\fr{U }{\widetilde {\cal V}}
d .{\tilde {\cal P}}\rt )^{-1}
?=?
\lt (
{\cal P}   
.d   .
\fr{U}{{\cal V}}\rt )^{-1}
\la{3t63tg3gwerf}
\ee
To prove this we note that the left side of
(\ref{3t63tg3gwerf}) is 
\be
\lt (\fr{U }{\widetilde {\cal V}}
d .{\tilde {\cal P}}\rt )^{-1}
=
 {\tilde {\cal P}}^{-1} . d^{-1} .
{\widetilde {\cal V}}
\ee
Use
\be
{\tilde {\cal P}}
=
\fr{1}{X} \lt (\fr{U}{ {\cal V} - d^{\dag} .\fr{U}{{\widetilde {\cal V}}} .d  } \rt)
\ee
so that
\be
{\tilde {\cal P}}^{-1}
=
X \lt (  {\cal V} - d^{\dag} .\fr{U}{{\widetilde {\cal V}}} .d   \rt)
\ee
So the left side of
(\ref{3t63tg3gwerf}) is 
\be
 X \lt (  {\cal V} - d^{\dag} .\fr{U}{{\widetilde {\cal V}}} .d   \rt) . d^{-1} .
{\widetilde {\cal V}}
=
 X \lt (  {\cal V} d^{-1} .
{\widetilde {\cal V}} - d^{\dag}  \rt )
\la{fgqrwgwerg}
\ee

The right side of
(\ref{3t63tg3gwerf}) is 
\be
\lt (
{\cal P}   
.d   .
\fr{U}{{\cal V}}\rt )^{-1}
=
{{\cal V}}.d^{-1} . {\cal P}^{-1}   
\ee
Now use
\be
{  {\cal P}}
=
\fr{1}{X} \lt (\fr{U}{\widetilde {\cal V} - d . \fr{U}{{ {\cal V}}} .d^{\dag}  } \rt)
\ee
so that
\be
{  {\cal P}}^{-1}
=
X \lt  ( {\widetilde {\cal V}} - d . \fr{U}{{ {\cal V}}} .d^{\dag}   \rt )
\ee
and hence the right side of
(\ref{3t63tg3gwerf}) is 
\be
{{\cal V}}.d^{-1} . X \lt  ( {\widetilde {\cal V}} - d . \fr{U}{{ {\cal V}}} .d^{\dag}   \rt )
=
X \lt ( {{\cal V}}.d^{-1}    {\widetilde {\cal V}} - d^{\dag}   \rt )
\ee
which is equal to the left side in (\ref{fgqrwgwerg})and so we have proved that
\be
{b}_{31}^{\dag}
=-
b_{13}
\ee
Along the way we have proved the identity
\be
\fr{U }{\widetilde {\cal V}}
d .{\tilde {\cal P}}=
{\cal P}   
.d   .
\fr{U}{{\cal V}}
\la{gqretujytuyk1}
\ee
Its hermitian conjugate is the identity:
\be
 \fr{U}{{\cal V}}
. d^{\dag} 
 \\  . {\cal P} 
={\tilde {\cal P}}
d^{\dag} .\fr{U}{{\widetilde {\cal V}}}
\la{gqretujytuyk2}
\ee
These identities recurs often in the demonstrations of hermiticity for the bosonic matrix ${\cal K}^{-1}$.

\item{Verify that $b_{14}= b_{41}^{\dag}$ }

This appears in \ref{boseansatz} in the term
\be
m^2 \\
 {b_{14}}^{q_1 s_4} 
\ee
which has no derivative and that gives rise to the 
requirement
$b_{14}= b_{41}^{\dag}$.  We have
\be
b_{14}=
-2 X^2  
{\widetilde {\cal G}}
\\
g^{\dag}.{\cal R}^{-1}.r .
\\
  \fr{U}{{\widetilde {\cal V}}} .
d.  \\
 {\tilde {\cal P}} . r.{\tilde {\cal R}}^{-1}
\\
.g^{\dag} .  {\cal G}
\\
\ee
\be
b_{41}=- 2 X^2 
{\cal G}
 .
\\  g.{\tilde {\cal R}}^{-1}
.r^{\dag} 
  . \fr{U}{{\cal V}}
.\\
 d^{\dag} 
   . {\cal P}  . r^{\dag}.{\cal R}^{-1}.g \\
 .{\widetilde {\cal G}}
\ee

We have
\be
{b}_{41}^{\dag}
=
-2 X^2  
{\widetilde {\cal G}}
\\
.g^{\dag}.{\cal R}^{-1}.r .
\\
{\cal P} .
d.\fr{U}{{\cal V}}
. r.{\tilde {\cal R}}^{-1}
\\
.g^{\dag} .  {\cal G}
\ee
so that the identity 
\be
{b}_{41}^{\dag}
=
{b}_{14}
\ee
will be true if and only if
\be
{b}_{14}=
  \fr{U}{{\widetilde {\cal V}}} .
d.  \\
 {\tilde {\cal P}} ={\cal P} .
d.\fr{U}{{\cal V}}
\ee
This was proved in 
equation (\ref{gqretujytuyk1}), so it follows immediately that
\be
{b}_{41}^{\dag}
=
{b}_{14}
\ee

\item{Verify that $b_{23}= b_{32}^{\dag}$}
\be
b_{232}=  \fr{1}{2 }
 \fr{U }{{\widetilde {\cal V}}}
d .{\tilde {\cal P}}
\\
-   \fr{1}{2 X}
d. {\tilde {\cal D}}
\\
\ee
\be
b_{233}=d. {\tilde {\cal D}}
\ee

\be
b_{322}=
 \fr{1}{2}\fr{U}{{ {\cal V}}} .
d^{\dag}.
{\cal P}
\\
-  
 \fr{1}{2 X}  
. d^{\dag} . {\cal D}
\\
\ee
\be
b_{323}=
- d^{\dag}. {{\cal D}}
\ee
We will have
\be
b_{322}^{\dag}= b_{232} 
\ee
if and only if
\be
\fr{U}{{\widetilde {\cal V}}} .
d.  \\
 {\tilde {\cal P}}=
{\cal P}
.\\
 d 
\fr{U}{{\cal V}}
   .  \ee
We have
\be
b_{233}^{\dag}= 
b_{323}
\ee
by the usual manipulation:
\be
d. {\tilde {\cal D}}=  {{\cal D}}.d
\ee
The identity:
\be
  d^{\dag} . {\cal D} 
=  {\widetilde{\cal D}} .d^{\dag}
\ee
 is obvious when one recalls the definitions 
\be 
{\cal D} = \fr{U}{ X^2 U - d . d^{\dag}}
\ee
\be 
{\widetilde {\cal D}} = \fr{U}{ X^2 U -   d^{\dag} . d}
\ee

\item{Verify that $b_{42}=- b_{24}^{\dag}$}
This appears in equation (\ref{boseansatz}) with one derivative, hence we expect:
$b_{42}= -b_{24}^{\dag}$
We have
\be
b_{24}=
 X \fr{U}{{\widetilde {\cal V}}} .
\\
d.   {\tilde {\cal P}} . 
\\
r.{\tilde {\cal R}}^{-1}
\\
.g^{\dag} .  {\cal G}
\ee

\be
b_{42}=- X {\cal G}.
g.{\tilde {\cal R}}^{-1}
\\
.r^{\dag}
.
 \fr{U}{{ {\cal V}}} .
\\
d^{\dag}.
{\cal P}
\ee

With the usual identity 
\be
\fr{U}{{\widetilde {\cal V}}} .
d.  \\
 {\tilde {\cal P}}=
{\cal P}
.\\
 d 
\fr{U}{{\cal V}}
   .  \ee
we see this is true.

\item{Verify that $b_{34}= -b_{43}^{\dag}$}
This appears in equation (\ref{boseansatz}) with one derivative, hence we expect:
$b_{34}= -b_{43}^{\dag}$
We have
\be
b_{34}=
- X {\tilde {\cal P}} . 
\\
r.{\tilde {\cal R}}^{-1}
\\
.g^{\dag} .  {\cal G}
\ee
\be
b_{43}=
X {\cal G}.g
\\
.{\tilde {\cal R}}^{-1}
.r^{\dag} 
\\.
 {\tilde {\cal P}} 
\ee
This is clearly true.
\een

\subsection{Special case when all relevant matrices commute for the Bosons--there are no poles except at the zeros of certain polynomials }
\la{notationappendixcommute}

One can verify that there are no poles in any of the terms of $b_{ij}$ except at the zeros  of the following polynomials:
\be
X^2 [X (U-R)^2  +   G]^2 - [X (U-R^2) + G]^2 D  =0
\ee
and  
\be
X^2 - D  =0
\ee

We shall examine some examples now. 

\subsubsection{Matrices that arise for the Bosons for the commuting case}

For the commuting case we find that:
\be 
{\cal D} = {\widetilde {\cal D}} =\fr{U}{ X^2 U - D}
\ee
\be
{  {\cal G}} = {\widetilde {\cal G}} = \fr{U}{( X  {{\cal R}}_{-} 
+  
g .   {\tilde {\cal R}}^{-1}
  . g^{\dag} )}
\Ra
\fr{(U+R)}{[ X  (U-R^2)+  G]}
\ee
\be 
{{\cal V}}=
{\widetilde {\cal V}}=  \fr{X [X (U-R)^2  +   G]
}{[X (U-R^2) + G]}
\ee
\be
{  {\cal P}}
=
{\tilde {\cal P}}
=
\lt (\fr{     [X (U-R^2) + G] [X (U-R)^2  +   G]
}{X^2 [X (U-R)^2  +   G]^2 - [X (U-R^2) + G]^2 D  } \rt)
\ee

\subsubsection{Verify that $b_{11}$ has only the usual poles}
\be
b_{11} =
- {\widetilde {\cal G}} . 
\\  + 2 X^2 {\widetilde {\cal G}} \\
.  
g^{\dag}.{\cal R}^{-1}.r .
\\ {\cal P}  . r^{\dag}\\.{\cal R}^{-1}.g  \\.{\widetilde {\cal G}}
\ee
This is clearly hermitian and is related  to $b_{44}$ by the tilde operation:
\be b_{44}=-
{\cal G}
\\
+
2 X^2 {\cal G}
\\. g.{\tilde {\cal R}}^{-1}
.r^{\dag}. 
\\  {\tilde {\cal P}} . r.{\tilde {\cal R}}^{-1}
\\.g^{\dag} .  {\cal G}
\ee
We need to take the commuting version of this and then add it up. 

We find that 
\be
b_{11}   =  
\lt (\fr{  \lt\{
-(U-R)X^2 
  [X (U-R)^2  
 +   G]   
 +  (U+R) D [X (U-R^2) + G]  
\rt \}
}{X^2 [X (U-R)^2  +   G]^2 - [X (U-R^2) + G]^2 D  } \rt)
\ee1

So we see that the only denominator left is the usual one.

\subsubsection{Case of $b_{322}$}

Consider the expression.  \be
b_{322} = \fr{1}{2}\fr{U}{{ {\cal V}}} .
d^{\dag}.
{\cal P}
-  
 \fr{1}{2 X}  
. d^{\dag} . {\cal D}
\la{qtrggeth}
\ee
As a further example, we want to show that there are no poles in this expression other than the ones mentioned above.  For commuting matrices with $d=d^{\dag}$ we find:
\be
b_{322} =  
\eb
\fr{d X }{ X^2 U - D}\lt (\fr{    
   [X (U-R^2) + G]^2
-      
 [X (U-R)^2  +   G]^2    }{X^2 [X (U-R)^2  +   G]^2 - [X (U-R^2) + G]^2 D  } \rt)
\ee
This demonstrates that there are no poles other than the usual ones.  In particular there is no $\fr{1}{X}$ in the sum, although it appears in the starting expression (\ref{qtrggeth}).

\section{Remarks about the parameter $R$, and the magnitude of gauge and supersymmetry breaking as seen in the effective action}

\la{gergrijergerioj}

We note that there are matrix parameters $r_{pq}$ and $R_{p}^{\;\;s}= r_{pq}
\ov r^{ q s}$ in the theory.  The matrix $r_{pq}$ first makes its appearance in the Superpotential for the Standard Model.  
The matrix $r_{pq}$   naturally gives rise to the hermitian positive semi-definite matrix 
\be
R_{p}^{\;\;s}= r_{pq}
\ov r^{ q s}
\ee

The matrix  $r_{pq}$ occurs again when the dotspinors  are incorporated into the new BRS transformations for the related dotspinor fields in subsections
\ref{leftleptonalgebra} and \ref{rightleptonalgebra}, and in the related actions in subsection \ref{fergergegeAction}.

These two uses of the matrices  $r_{pq}$  are not quite the same however.  They differ by an overall factor, and something more too.

The difference arise from the necessity of starting with 
the equation:
\be
\d_{\rm SS} \ov \w_{\a}^{ q}=
\ov r^{ q  p }
\og
\ovv^2
 m^2 C_{\a} \; 
\ov R_{ p }
\ee
and transforming this to the expression 
\be
\d_{\rm GSB} \ov \w_{R \a}^{ q}=
\ov r^{ q  p }
C_{\a} \; 
\ov A_{R  p }
\ee
This latter expression occurs in  the component version of the cybersusy algebra in \ref{rightleptonalgebra}.

Again the matrix 
\be
R_{p}^{\;\;s}= r_{pq}
\ov r^{ q s}
\ee
arises in the new context.  But its range is restrained:
\be
0\leq R < U
\ee
This restraint arises if we want to prevent the kinetic term of various fields from getting the wrong sign.  
For example in the kinetic matrix 
 for the bosons \ref{matrixforbosons}, the upper left  entry contains
\be
- \D
\\
( \d -
 {\tilde R})_{\;\;q_1}^{ p _1}
\ee
and if 
\be
(\d -
 {\tilde R})_{\;\;q_1}^{ p _1} 
\ee
changes sign, that looks rather serious.

But the range of the original matrix $R_{p}^{\;\;s}$ was not restrained. What is happening here?

There are the following  differences:
\ben
\item 
a difference of the names of the fields;  
\item 
a difference of  the mass dimension of the fields;
\item
we do not attempt to write down an action for the composite fields, and we do write down an action for the effective fields;
\item
 a difference of the following factor:
\be
\og
\ovv^2
 m^2 ;
\ee
\item
 a difference of normalization of the fields.
We note that:
\be
v^2 = \fr{g_{\rm J}}{g}
\ee 

So the factor is
\be
\og
\ovv^2
 m^2 
=
\og
\fr{\og_{\rm J}}{\og}
 m^2 
=
 \og_{\rm J} 
 m^2 
\ee

We choose to normalize the new fields 
$\w_{R \a}^{ q}$
and $
\ov A_{R  p }
$ in such a way that their kinetic terms are standardized.  But what does this mean about the normalization of the new
parametric matrix $\ov r^{ q  p }$?
\item
The parameter $m$ in the SSM and the effective action  are also not the same, since any possible connection between them would presumably be affected by an incalculable renormalization constant relating to bound states.
\een

It appears that there is a curious feature in  that the overall signs of the terms 
\ref{rqewfwfwefweL} and
 \ref{rqewfwfwefweR}  are not determined by the positivity of the Hamiltonian, because neither sign guarantees positivity. In fact the Hamiltonian analysis is problematic because there are more than two time derivatives in some places in the action (the $\w \D \pa \ov \w$ part). This needs further thought.
It appears that the result for the other sign might be more or less the same, but it would be nice to know.


\tableofcontents

\end{document}